\documentclass[aps,prx,twocolumn,footinbib,superscriptaddress,floatfix]{revtex4-2}
\usepackage{graphicx}
\usepackage{indentfirst}
\usepackage{physics}
\usepackage{braket}
\usepackage{float}
\usepackage{mathtools}
\usepackage{epstopdf} 
\usepackage{footnote}
\usepackage{esint}
\usepackage{comment}
\usepackage{color}
\usepackage[T1]{fontenc}
\usepackage{subfigure}
\usepackage{amsfonts}
\usepackage{footmisc}
\usepackage{scrextend}
\usepackage{multirow}
\usepackage[hyperfootnotes=false]{hyperref}
\usepackage[acronym]{glossaries}
\usepackage[english]{babel}
\usepackage{url}
\usepackage{bm}
\usepackage{algpseudocode}
\usepackage{svg}
\usepackage{etoolbox}
\usepackage{enumerate}
\definecolor{darkblue}{rgb}{0,0,0.5}
\hypersetup{
    colorlinks=true,
    linkcolor=black,
    filecolor=blue,
    citecolor=darkblue,  
    urlcolor=black,
}

\apptocmd{\sloppy}{\hbadness 9999\relax}{}{}

\newcommand{\calL}{{\cal L}}

\newcommand{\calT}{{\cal T}}

\newcommand{\calH}{{\cal H}}

\newcommand{\1}{^{(1)}}

\def\be{\begin{equation}}
\def\ee{\end{equation}}
\def\ba{\begin{eqnarray}}
\def\ea{\end{eqnarray}}
\def\bal{\begin{equation}\begin{aligned}}
\def\eal{\end{aligned}\end{equation}}

\usepackage[normalem]{ulem}

\begin{document}

\title{Quantum-enhanced radio-frequency photonic distributed imaging  }
\author{Haowei Shi}
\affiliation{
Ming Hsieh Department of Electrical and Computer Engineering, University of Southern California, Los
Angeles, California 90089, USA}

\author{Christopher M. Jones}
\affiliation{Halliburton Technology, 3000 N Sam Houston Pkwy E, Houston, Texas 77032, USA}

\author{Mengjie Yu}
\affiliation{
Ming Hsieh Department of Electrical and Computer Engineering, University of Southern California, Los
Angeles, California 90089, USA}

\author{Zheshen Zhang}
\affiliation{
Department of Electrical Engineering and Computer Science,
University of Michigan, Ann Arbor, MI 48109, USA
}

\author{Quntao Zhuang}
\email{qzhuang@usc.edu}

\affiliation{
Ming Hsieh Department of Electrical and Computer Engineering, University of Southern California, Los
Angeles, California 90089, USA}
\affiliation{Department of Physics and Astronomy, University of Southern California, Los
Angeles, California 90089, USA}

\begin{abstract}
Quantum physics has brought enhanced capability in various sensing applications. Despite challenges from noise and loss in the radio-frequency (RF) domain, [Phys. Rev. Lett. 124, 150502 (2020)] demonstrates a route for enhanced RF-receiver empowered by quantum squeezing and entanglement. In this work, we further explore the quantum advantage of imaging in the weak coupling scenario of the RF-photonic receiver. The proposed imaging receiver applies transducer to upconvert the RF signal to optical to enable high-efficiency connection via low-loss fiber networks. The efficient connection therefore increases the synthetic aperture and improves the resolution of the distributed imaging system. To overcome the challenge from low transduction efficiency in existing devices limited by weak photon interaction, we propose the use of squeezed-state optical sources to suppress the noise. We numerically evaluate the quantum advantage in synthetic aperture radar imaging, where the images are generated from the standard resolution test chart via a Gaussian point spread function with added Gaussian noise. We apply the Wiener filter on the images to restore the objects and find that stronger squeezing significantly improves the quality of the restored image. Our findings push quantum squeezing advantage to real-world applications.
    
\end{abstract}

\maketitle

\section{Introduction}
Quantum effects such as entanglement and squeezing are not only conceptually surprising, but also can enhance the performance of information processing tasks, including sensing applications~\cite{giovannetti2006,pirandola2018advances,lawrie2019quantum,zhang2021dqs}. A major paradigm in achieving quantum advantage in sensing is by deploying quantum probes that actively interacts with the sample and then measuring the returned probes with quantum measurement. Prominent examples include Laser Interferometer Gravitational-Wave Observatory (LIGO)~\cite{abadie2011gravitational,aasi2013enhanced,tse2019quantum} and Haloscope At Yale Sensitive To Axion CDM (HAYSTAC)~\cite{backes2021} for dark matter detection, where squeezed light are injected to improve the signal-to-noise (SNR) ratio. Such applications require well-controlled experimental conditions to minimize loss and noise such that quantum advantage can be achieved. In other scenarios such as lidar detection~\cite{giovannetti2001quantum,zhuang2017entanglement,huang2021quantum,reichert2022quantum,reichert2023heisenberg}, loss is inevitable and the possible quantum advantage reduces when loss increases.

In the radio-frequency (RF) region, the abundant background noise makes quantum sensing advantages challenging, as the sensing process destroys quantum effects such as squeezing and entanglement. Despite quantum illumination~\cite{tan2008quantum,zhang2015,assouly2023quantum,zhuang2022} providing surprising advantage in target detection tasks, its low-power operating region makes it hard to compete with classical RF sensors~\cite{shapiro2020quantum}. Indeed, in the RF frequency region, practical quantum sensing advantage with active probes seems challenging. Another route towards practical sensing advantage focuses on the measurement. In the RF region, Ref.~\cite{xia2020demonstration} provides a paradigm of practical quantum advantage in RF-photonic sensing scenarios. Although the probe state is classical, reaching the classical sensing limit with existing device can be challenging, especially when a RF antenna array is involved due to the large wavelength. 

RF waves at frequency below conventional GHz radar band enjoy a smaller power attenuation rate in subterranean sensing~\cite{korpisalo2014characterization}, thus drastically higher SNR at a fixed distance. However, their long wavelengths lead to severe diffraction blurs, which require unrealistic aperture sizes at the kilometer level.
In the regime of long wavelength above centimeters, synthetic aperture radars (SARs) ~\cite{franceschetti2018synthetic,chapin2012airmoss} are designed to ease the requirement on aperture size. It uses a moving antenna or an antenna array \cite{krieger2007multidimensional} to shine a coherent signal beam on an object, and collect the returned signal pattern coherently to simulate a significantly larger synthetic aperture size. To circumvent the resolution limit determined by the bandwidth and the synthetic aperture size (limited by the RF cable loss), RF-photonic radars and SARs have been proposed~\cite{ghelfi2014fully,zhang2017photonics} to generate and process high-frequency high-bandwidth RF signals. Also, for the interferometric SAR~\cite{dong2020microwave}, RF-photonics enable the low-loss optical fiber network connecting the spatially distributed antennas. As it is challenging to increase the transduction efficiency to unit, the SNR is extremely low. While diffraction blur can be improved by the larger aperture size enabled by photonics, it is also important to reduce the detector noise, as the imaging reconstruction problem is highly sensitive to noise~\cite{frieden1979image,den1997resolution,frieden1967band,frieden1972restoring,frieden1979image}.

In this work, we generalize the paradigm of quantum RF-photonic sensing to tackle imaging problem, with particular applications in subterranean sensing. To enhance the SNR of RF-photonic imaging system, we propose squeezing the noise of the optical source before the upconversion, as the optical noise dominates at the low transduction efficiency limit.
We numerically simulate the squeezing-enhanced image reconstruction for general object pattern in the low SNR regime, where the single-mode squeezing techniques have been well established~\cite{tse2019quantum,aasi2013enhanced}, and the SNR is commonly low due to the low efficiency of transduction needed in low-frequency radars. Our results show a clear advantage from squeezing in the images, also quantified by the commonly adopted peak SNR (PSNR) of the images~\cite{salomon2002data}. At high loss scenario, squeezing provides one-order-of-magnitude resolution enhancement, thanks to the nonlinear threshold phenomena in non-asymptotic sensing region, with $10$dB of squeezing.




\section{Protocol}

An SAR uses a moving antenna or an orthogonally-coded antenna array as the transceiver to coherently shine a large beam of temporally chirped pulse signal on a lossy object, and coherently collect the returned signal pattern, which is chirped in both the temporal and spatial domain due to the Fresnel diffraction. With the collected return, it then applies a matched filter to reconstruct the object pattern $f(x,y,z)\equiv \sqrt{\kappa(x,y,z)}e^{i\theta(x,y,z)}$, where $\kappa$ is the power transmissivity (or reflectivity), $\theta$ is the phase shift, $x,y$ are the transverse dimensions, and $z$ is the range dimension. In this paper, we ignore the range dimension $z$ and consider only the transverse dimension $x,y$. 

\begin{figure}[t]
    \centering
    \includegraphics[width=\linewidth]{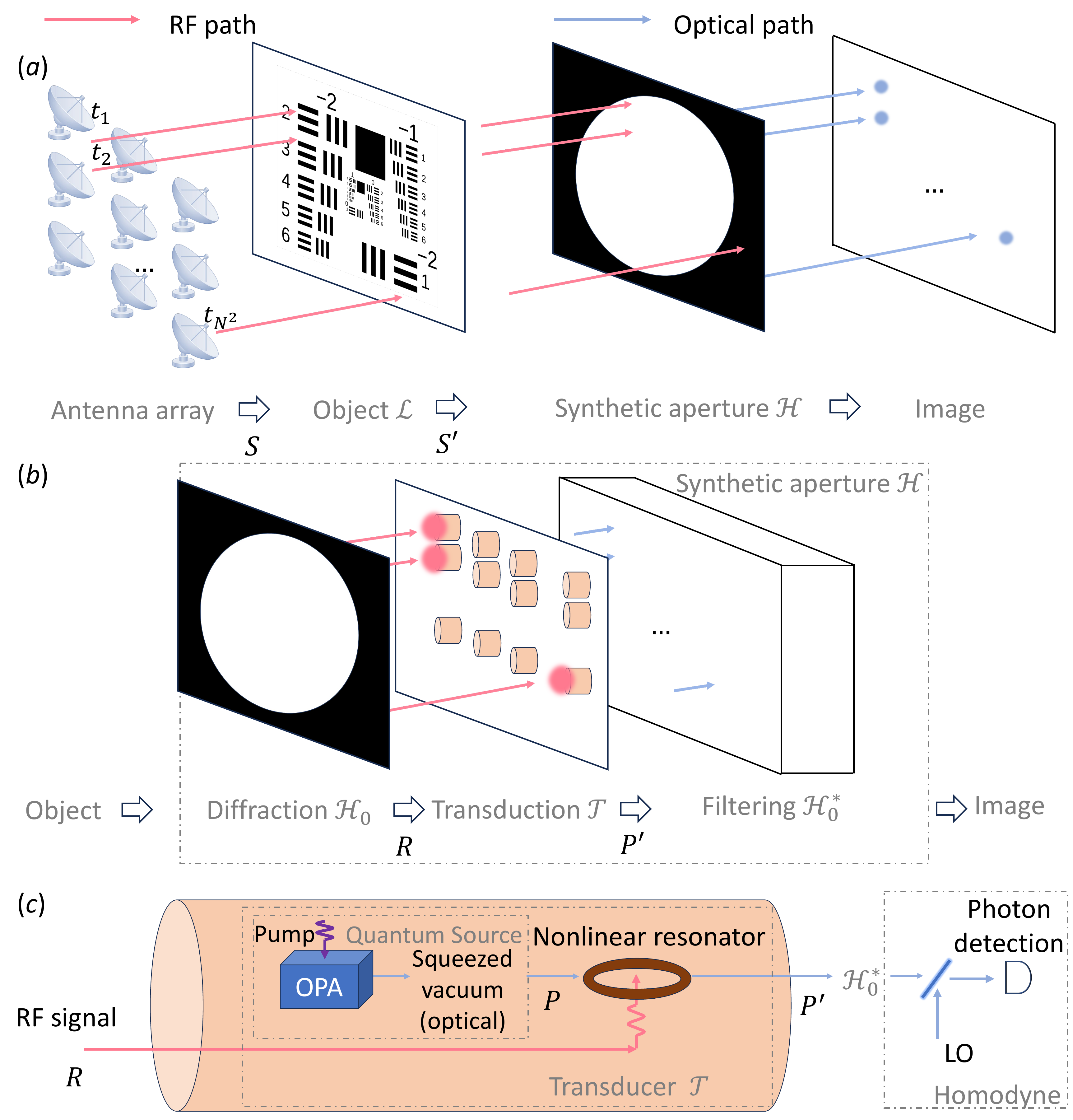}
    \caption{(a)Overview of the entanglement-enhanced radio-frequency(RF) photonics imaging protocol. Orange cylinders: receivers, each consisting of a transducer and a photon detector. The antenna components in the $N\times N$ antenna array operate at orthogonal modes (e.g. temporal or spatial multiplexing) individually. It sends the signal $S$ to the sample, which is modulated to $S'$. Then the object $S'$ is imaged to the image via an effective synthetic aperture. (b)The mechanism of the synthetic aperture $\calH$. The object first propagates to the transducer array subject to diffraction $\calH_0$, then gets transduced to the optical probe, finally jointly processed by the matched filter $\calH_0^*$. (c) Schematic of the transducers and the measurement at the image. Each transducer embeds an RF return $R$ into the optical probe $P$, then the output $P'$ is coherently modulated by the matched filter $\calH_0^*$, and finally homodyne measured. LO: local oscillator. 
    }
    \label{fig:experiment}
\end{figure}

An overview of our protocol setup is shown in Fig.~\ref{fig:experiment}. After traveling through the object $f(x,y)$ (e.g. the USAF 1951 resolution test chart in this figure), the signal beam propagates to the image plane, each point of $f(x,y)$ is blurred by the point spread function (PSF) $h(x,y)$.
Signal chirping widens the power spectrum and narrows the PSF that simulates a synthetic aperture larger than the physical antenna aperture. For the sidelooking SAR, a spatial matched filter, i.e., the chirp compression process, narrows the PSF $h(x,y)$ equivalent to that associated with a synthetic aperture, e.g., of size equal to twice of the beam size in a strip-mapping SAR~\cite{munson1989signal}. As a result, the waist size of PSF is $w_0=D/2$, where $D$ is the physical antenna aperture~\cite{munson1989signal}. To implement the matched filter coherently, we first upconvert the RF signals to the optical domain, and then combine the optical signals using an extremely low-loss optical information processing. For the upconversion, we use an electro-optic modulator (EOM) to couple the RF signal to the sideband mode of a strong optical pump.

\begin{figure*}
    \centering
    \includegraphics[width=\linewidth]{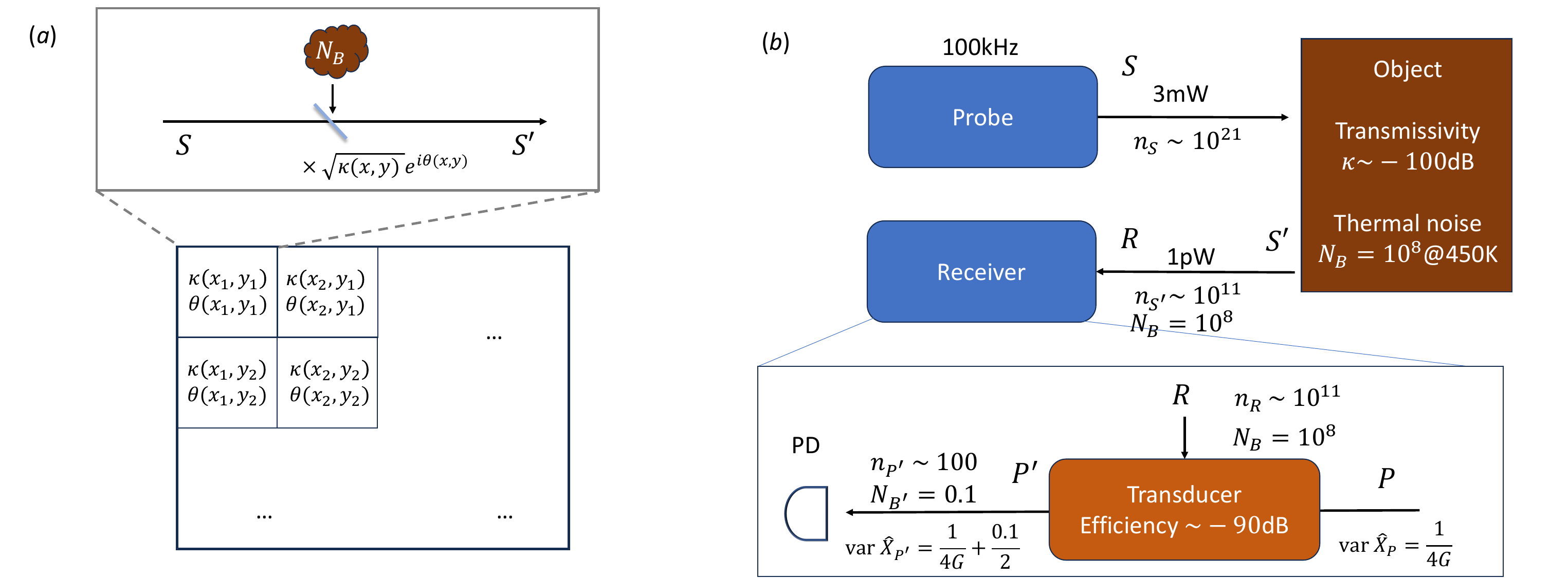}
    \caption{(a) Quantum model of the object. (b) A practical parameter setup. The detected thermal photon number is below shot noise level.}
    \label{fig:schematic}
\end{figure*}

Below we elaborate the protocol step by step, starting from the antenna to the target, and finally to the transduction and measurement at the receiver side. A schematic of the model for theory analyses is shown in Fig.~\ref{fig:schematic}, where the target induces loss and noise from thermal background and the transduction system has limited efficiency and additional shot noise. Our major goal is to suppress the shot noise at the transduction step, while the other noises are intrinsic and unavoidable.

\emph{Antenna setup.---}
We consider a uniformly distributed antenna array, each of aperture size $D$. We will consider the effect of the sampling rate, i.e., the antenna spacing, in Appendix~\ref{sec:digitalization}.
A pulse-compressed source of time duration $T$ and bandwidth $B$ yields $M=BT$ time modes, equivalent to $M$ independent probes. 
To constrain the energy consumption, we fix the total source radiation power to $P_S$. Then the total (whole-image) source photon number per time mode is
\be 
n_S= \frac{P_S T}{\hbar \omega \cdot M}=\frac{P_S }{\hbar \omega  B},
\ee
where $\hbar$ is the Planck constant, $\omega$ is the frequency of the source in unit of radian per second. For radiation power $P_{S}=3$mW operating at $\omega=$100kHz, $n_S\simeq 10^{21}$ with $B=\omega/2=50$kHz. We note that at the long wavelength limit, the antenna is in the short dipole regime where the total power consumption rapidly increases with the wavelength: $P_{\rm tot}\propto \lambda^2 P_{S}$~\cite{staelin2011electromagnetics}.

\emph{Field quantization.---}
To enable the treatment of quantum effects such as squeezing, we briefly introduce the theory tools of quantum optics. The complex electromagnetic field is quantized as an annihilation operator $\hat a=\hat X+i\hat Y$, which consists of two orthogonal quadratures in the phase space: $\hat X$ proportional to the electric field and $\hat Y$ proportional to the magnetic field. The power consumption $P$ is proportional to the photon number $\hat a^\dagger \hat a$: $P=\hbar \omega \hat a^\dagger \hat a/T$. Field annihilation operators satisfy the canonical commutation relation $[\hat a(\bm x), \hat a^\dagger(\bm x') ]=\delta(\bm x-\bm x'), [\hat X(\bm x),\hat Y(\bm x')]=i/2\cdot \delta(\bm x-\bm x')$, where $\delta(\cdot)$ is Dirac function. The commutation relation yields the famous Heisenberg uncertainty principle: consider a narrow spatial mode at arbitrary position $\bm x$ such that $[\hat X,\hat Y]=i/2$, one can show that the fluctuation $\expval{\Delta \hat X^2}\cdot \expval{\Delta \hat Y^2}\ge |[\hat X,\hat Y]|^2/4=1/16$. For a vacuum state free from any thermal and technical noise, it has the quadrature fluctuations $\expval{\Delta \hat X^2}=\expval{\Delta \hat Y^2}=1/4$ where $\Delta \hat X=\hat X-\expval{\hat X}$, which is also known as the standard quantum limit.  On the other hand, we propose to squeeze the noise $\expval{\Delta \hat X^2}$ on one quadrature via degenerate parametric amplifiers, at the cost of amplifying the other quadrature $\expval{\Delta \hat Y^2}$. Below we briefly introduce the noise squeezing mechanism.

In our quantum enhanced protocol, the light probe is squeezed using an optical parametric amplifier, i.e., the quadrature noise is suppressed beyond the standard quantum limit, thus the detected SNR is improved. To enable the analyses, we provide some background of squeezing. An spontaneous parametric downconversion (SPDC) using periodic poled lithium niobate (PPLN) can generate squeezed light~\cite{hao2021entanglement,hao2022demonstration}, which reduces the fluctuation on one quadrature and inevitably amplifies the fluctuation on the other due to the uncertainty principle. Without loss of generality, let the squeezed quadrature be $\hat X$, then $\expval{\Delta \hat X^2}=1/4G$, $\expval{\Delta \hat Y^2}=G/4$ with squeezing gain $G$. 
In practical experiments, effective gain $G$ at the detector side can already be as high as $10^{1.5}$ (15dB)~\cite{vahlbruch2016}.

\emph{Object $\calL$.---}
As shown in Fig.~\ref{fig:schematic}(a), the source photons are shined on the object to be measured. We formulate the object as a spatial array of bosonic thermal-loss channels denoted as $\calL_{\kappa(x,y),N_B,\theta(x,y)}$ that maps the signal field annihilation operator $\hat a_S\to \hat a_S'$, with a constant additive thermal background of dark photon count number $N_B\equiv \expval{\hat a_S^{\prime \dagger}\hat a_S'}$ when the input $\hat a_S$ is vacuum and a spatial distribution of power transmissivity $\kappa(x,y)$ and phase shift $e^{i\theta(x,y)}$. In the phase space of $\hat X, \hat Y$, the thermal background invokes an additive circular-symmetric two-dimensional Gaussian noise of variance $\expval{\Delta \hat X^2}=\expval{\Delta \hat Y^2}=(1+2N_B)/4$. 
We can model the input-output relation for each pixel at location $(x,y)$ as a bosonic Gaussian quantum channel~\cite{weedbrook2012gaussian} 
\bal 
~&\calL^{S\to S'}_{\kappa(x,y),N_B,\theta(x,y)}: \hat a_S(x,y)\to \hat a_S'(x,y)\,,\\
&\hat a_S'(x,y)= \sqrt{\kappa(x,y)}e^{i\theta(x,y)}\hat a_S(x,y) +\sqrt{1-\kappa(x,y)} \hat e(x,y)\,,
\eal
where the environment is a Gaussian white noise of thermal photon number $\expval{\hat e^\dagger (x',y')\hat e(x,y)}=N_B/[1-\kappa(x,y)]\delta(x-x',y-y')$.
A typical parameter setup of the protocol is shown in Fig.~\ref{fig:schematic}(b) as an example.

\emph{Imaging system $\calH$.---}
For a strip-map mode SAR~\cite{munson1989signal}, the waist size $w_0=D/2$ of the point spread function is 
half of the the physical antenna aperture size $D$.
The PSF can be modeled as a Gaussian form, 
\be 
h(x,y)=\frac{1}{\sqrt{\pi w_0^2} }\cdot \exp{-\frac{x^2+y^2}{2w_0^2 }}\,.
\ee
Consider an arbitrary object field $\hat a_S'(x,y)$ and include the transduction efficiency $\eta$, the image system forms the map
\bal 
~&\calH^{S'P\to P'}: \Big(\hat a_S'(x',y'), \hat a_P(x,y)\Big) \to \hat a_{P'}(x,y)\,,\\
&\hat a_{P'}(x,y)= \sqrt{\eta} \iint dx' dy' \hat a_S'(x',y') h(x-x',y-y')\\
&\qquad \qquad \quad +\sqrt{1-\eta}\hat a_P(x,y).
\eal
We denote the convolution above as a spatial map $\calH_{w_0}$. 

Below, we summarize the physical mechanism of the imaging system $\calH$, as illustrated in Fig.~\ref{fig:experiment}(b). 
Overall, $\calH^{S'P\to P'}=\calH_0^{*P'\to P'}\circ\calT^{RP\to P'}\circ\calH_0^{S'\to R}$, which consists of three stages: diffraction $\calH_0$ from the return $S'$ before the aperture to $R$ before the transduction, transduction $\calT$ from the in-aperture return $R$ onto the optical readout probe $P'$ (where the initial optical probe is $P$), and the chirp compression $\calH_0^*$. 
The diffraction $\calH_0$ is a Fresnel spatial convolution that characterizes the transverse blur during the free-space paraxial propagation of the RF beam for range $z$, while $\calH_0^*$ is simply the matched filter for $\calH_0$~\cite{munson1989signal}. 

The transducer schematic is shown in Fig.~\ref{fig:experiment} (c). A transducer invokes a nonlinear interaction between the RF signal $R$ and the optical probe $P$ in a nonlinear medium, e.g. lithium niobate. Overall the transducer forms a two-mode interaction with the RF output port discarded: $RP\to P'$. The input-output relation from the RF return to the optical probe mode can be modeled as a beamsplitter interaction $\calT^{RP\to P'}$ as
\be 
\hat a_{P'}=\sqrt{\eta}\hat a_R+\sqrt{1-\eta}\hat a_P
\ee
where $\hat a_P$ is the probe state which is engineered to be a desired quantum state with suppressed fluctuations. We see that the RF signal is cast on the optical probe output by power transmissivity $\eta$. In an electro-optic modulator, $\eta$ is limited by the optical pumping power $P_{in}$ and the coupling strength modelled by the half-wave voltage $V_\pi$ as $\eta=\frac{\pi^2}{4V_\pi^2}\frac{\omega_{e}}{\omega_o}Z\cdot P_{in}$, where $Z$ is the RF impedance (typically $50\Omega$), $\omega_{e/o}$ is the RF/optical frequency.  
The on-chip transduction efficiency $\eta$ is extremely small $\eta\ll 1$, typical state-of-the-art values are $\sim 10^{-5}$~\cite{holzgrafe2020cavity} even with the assistance of high-Q cavity.
In commercially available systems that are suitable for in-field deployment, the transduction efficiency is typically on the order of $\eta=10^{-9}$ (90dB) for $V_\pi=1 $V and $P_{in}=100$mW, $\omega_e=100$kHz, $\omega_o=192$THz. Denote the average of target transmissivity $\kappa(x,y)$ as $\overline \kappa\equiv\sum_{i=1}^{m_o}\sum_{j=1}^{n_o}\kappa(x_i,y_j)/m_o n_o$ (for $m_o\times n_o$ total pixels), the total information photon number from the RF signal transduced to the output probe mode is approximately $n_{P'}\equiv\eta \overline\kappa n_S$. 



Combining the object model $\calL$ and the imaging system $\calH$, we can summarize the overall input-output relation. Given a probing mode at transmitter $\hat a_S$ and the optical probe input $\hat a_P$, the output is
\be 
\hat a_{P'}= \calH_{w_0}^{S'P\to P'}\circ \calL_{\kappa(x,y),N_B,\theta(x,y)}^{S\to S'}( \hat a_S\otimes  \hat a_P)
\ee
where the thermal loss channel $\calL$ models the object pattern $f(x,y)=\sqrt{\kappa(x,y)}e^{i\theta(x,y)}$, the PSF convolution $\calH$ models the imaging system of the synthetic aperture assisted with the RF-photonic transduction. 

\emph{Measurement.---}
We consider a homodyne measurement on the quadrature $\hat X_{P'}= \Re\left(\hat a_{P'}\right)$ of the transducer output $P'$, as shown in Fig.~\ref{fig:experiment}(c). The homodyne measurement is fulfilled by coherently beating the input with a strong in-phase local oscillator (LO). The readout of $\hat X_{P'}$ is a random variable subject to Gaussian distribution. 

With an adaptive phase compensation $\theta_C(x,y)$, its mean $\expval{\hat X_{P'}}=\sqrt{\eta n_S} \Re \left(f(x,y)e^{i\theta_C(x,y)}\right)= \sqrt{\kappa(x,y)\eta n_S}\cos{[\theta(x,y)+\theta_C(x,y)]}$ carries the information about the object $f(x,y)$. For the transmissivity information, by adaptively choosing $\theta_C(x,y)\to -\theta(x,y)$, the induced signal differential $\partial_{\sqrt{\kappa}} \expval{\hat X_{P'}}(x,y)=\sqrt{\eta n_S}$ is proportional to $\expval{\hat X_{P'}}$. For the phase information, by adaptively choosing $\theta_C(x,y)\to -\theta(x,y)+\pi/2$, the induced signal differential $\partial_{\theta} \expval{\hat X_{P'}}(x,y)=\sqrt{\kappa(x,y)\eta n_S}$ is proportional to $\expval{\hat X_{P'}}$. Thus it is fair to define the signal difference in SNR as $\expval{\hat X_{P'}}$. 

The fundamental limit of the measurement variance is set by the squeezed vacuum state at the quantum probe system $P$. Given squeezing gain $G$, the quantum-limited quadrature fluctuation of squeezed vacuum with $\eta\to 1, N_B\to 0$ is $\text{var }(\hat X_{P'})=1/4G$. In practice, the noise in RF signal contributes to a small dark photon count $N_B'\equiv \eta N_B$ in the detection, which results in an extra variance. 
Combining the signal and noise, we can now derive the SNR of the system.
Consider the transmissivity average $\overline\kappa$. The homodyne measurement is characterized by the mean field $\expval{\hat X_{P'}}^2=\sqrt{n_{P'}}$, and the quadrature fluctuation is $\text{var }(\hat X_{P'})=[(1-\eta)/G+2N_{B}']/4$. The average whole-image SNR per-time-mode is
\be 
{\rm SNR}=\frac{\expval{\hat X_{P'}}^2}{\text{var }(\hat X_{P'})}=\frac{4\eta \overline\kappa n_S}{(1-\eta)/G+2\eta N_B}.
\label{eq:snr}
\ee

\begin{figure*}
    \centering
    \includegraphics[width=\linewidth]{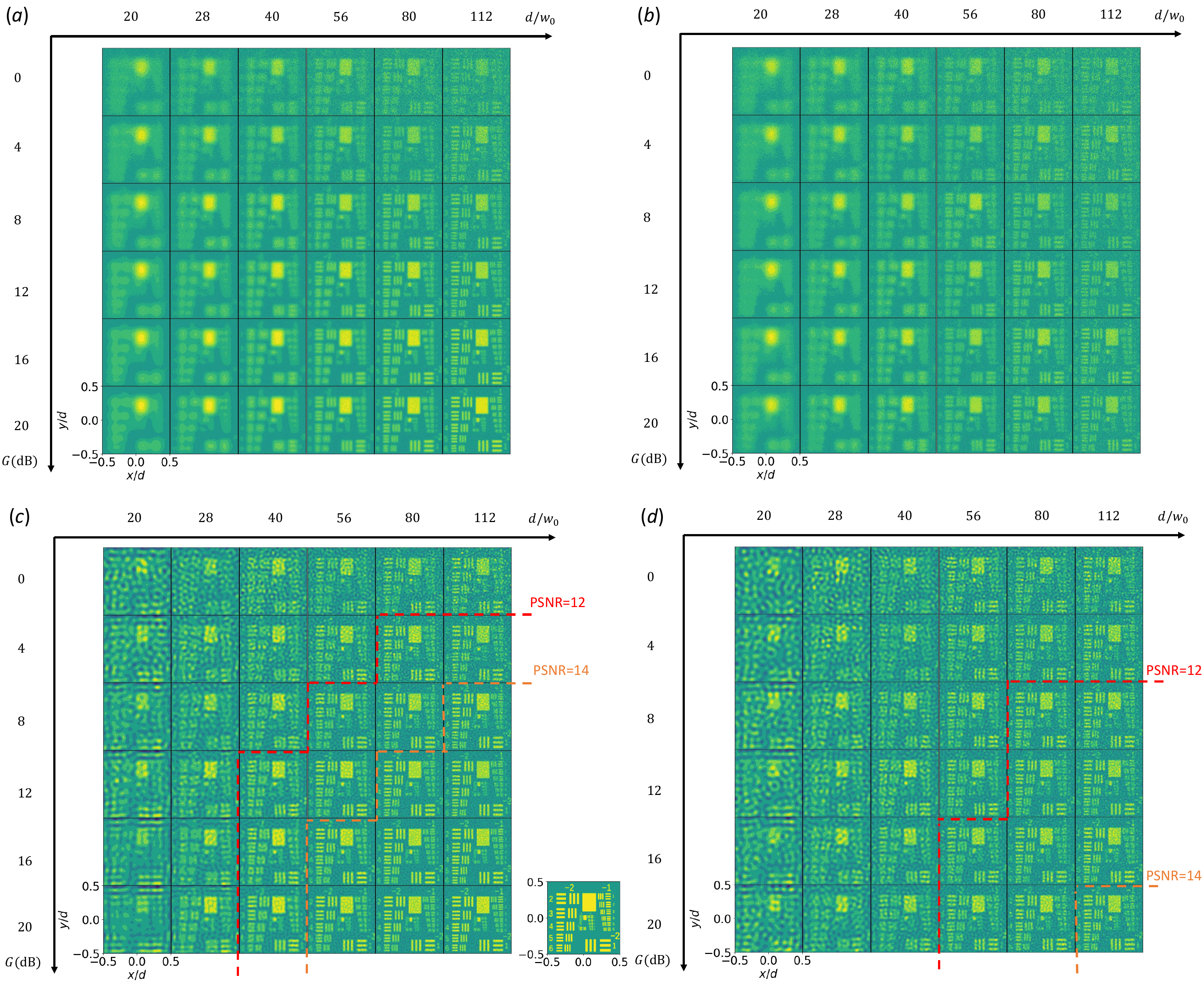}
        \caption{The raw blurred images (a)(b) of the USAF 1951 resolution test chart and Wiener-deconvolution reconstructions (c)(d) from the raw images. Noise setup: (a)(c) quantum-limited, transduced dark photon number $N_B^\prime\equiv \eta N_B=0$; (b)(d) thermal, $N_B^\prime=0.1$. Bottom-middle inset: the ground truth object, pixel dimensions $m_o=n_o=200$. 
        Total received photon number $n_{P'}=100$. Each subplot contains $6\times 6$ image reconstructions under various parameter setups: from left to right, object size $d$ increases relative to diffraction blur size $w_0$; from top to bottom, squeezing gain $G$ increases. The red and orange contours mark the contours of PSNR$=12$, below which the reconstructed image is not recognizable, and PSNR$=14$, above which the reconstructed image is well recognizable. Blurring is shown avoidable with higher squeezing gain $G$.  
        }
    \label{fig:reconstructed_6x6_USAF}
\end{figure*}

\emph{Image reconstruction.---}
To reconstruct the object, we apply the Wiener deconvolution~\cite{bertero2021introduction} to the image. We use the PSNR as the performance benchmark, which is proportional to the inverse of the mean square error of the reconstructed image $\tilde f(x,y)\equiv\sqrt{\tilde\kappa}(x,y)$ from the ground truth $ f(x,y)\equiv\sqrt{\kappa(x,y)}$:
\be 
{\rm PSNR}=10\log_{10}\frac{\max_{i,j}|f(x_i,y_j)|^2}{\sum_{i=1}^{m_o}\sum_{j=1}^{n_o}|\tilde f(x_i,y_j)-f(x_i,y_j)|^2/m_on_o}\,.
\label{eq:psnr}
\ee
Typically, PSNR on the order of 10-20 dB are considered as the minimum level to be a recognizable image, which is consistent with what we see in Fig.~\ref{fig:reconstructed_6x6_USAF}.

\section{Application setting}

In general, the proposed RF-photonics imaging applies to various scenarios where RF signals are used to probe the targets. A particular application of interest is the subterranean imaging that is of interest to geophysics, construction and petroleum engineering. In such applications, penetration depth and resolution are two important metrics of sensing systems. Electromagnetic waves propagate in subterranean substance with significant loss, especially at higher frequencies~\cite{korpisalo2014characterization}. While lower frequency (e.g., 100kHZ) transmits to larger depth, the longer wavelength requires a larger aperture for the antenna systems to produce acceptable resolution. Due to the large wavelength, the resolution of conventional imaging is unacceptable at low frequency and large wavelength (e.g. 300 meters for 100kHZ). The state-of-the-art low-frequency synthetic aperture radar operates at the P band, which can go down to $300$MHz. Considering the challenges in subterranean sensing, RF-photonic transducer allows low-loss coherent signal distribution and collection through fiber connection, thus increases the synthetic aperture size and improves resolution. 

The major challenge for RF-photonic sensing is the SNR~\cite{ghelfi2014fully}, as classical commercially available transducers suitable for large-scale field deployment typical transduction efficiency of 90 dB, severely lmiting the image quality. The quantum squeezing that is readily available at optical frequencies suppresses the vacuum noise from the huge loss, thus enable the distributed imaging even with a commercial low-efficiency transducer, and achieve much larger penetration depth affordably.

Here we consider the typical parameters subterranean geo-information detection: the resistive medium invokes electromagnetic transmissivity $\kappa(x,y)\simeq \overline\kappa=10^{-10}$, i.e. 100 dB loss; thermal photon number $N_B\simeq 10^8$ at $450$K (350$^\circ$F) for 100kHz carrier frequency; transduction efficiency for a commercial electrooptic transducer $\eta=10^{-9}$ (-90dB). We illustrate the parameters with their physical effect in our protocol in Fig.~\ref{fig:schematic}. For transmitter power $P_S=$0.01Watts (which requires ~1Watt of power for the entire transmitter to function), the total information photon number from the RF signal collected at the photon detector is $n_{P'}=\eta \overline n_S\simeq 100$. Meanwhile, the transduced background dark photon count number is $N_{B}'=\eta N_B=0.1$. At this weak SNR regime, squeezing is necessary to retrieve an acceptable resolution.

\section{Results}

With the parameters given in the previous section, now we proceed to perform numerical simulation on the imaging process. Note that the time mode number $M$ is a highly limited resource that are consumed not only for the transverse $xOy$ plane imaging, since the temporal profile of the $M$ signal modes can carry the range $z$ information which are not considered in this paper. Thus we take $M=1$ to benchmark the per-time-mode imaging performance for simplicity. If one adopts longer integration time to have large $M$, then one can further extend the penetration depth---achieve the same SNR at a smaller $\eta$ (see Eq.~\eqref{eq:snr}), while the quantum advantage we obtain in this section still holds at a larger penetration depth where SNR approaches the resolvable limit. 



As a test problem, we consider the USAF 1951 resolution test chart (see Fig.~\ref{fig:reconstructed_6x6_USAF}(a) bottom right) as the true target to design the $\kappa(x,y)$ distribution. We perform the simulation at different values of squeezing gain $G$.
As shown in Fig.~\ref{fig:reconstructed_6x6_USAF}, the resolution significantly improves as the quantum noise squeezing gain $G$ increases along the y-axis direction, especially when the diffraction blur size $w_0$ is negligible relative to the object size $d$. Thus $d\gg w_0$ is the scenario of our interest. By contrast, for small object size $d\lesssim 20 w_0$, we see a saturation for $G$, i.e. increasing $G$ further does not improve PSNR. This is because in this case the structures of the object components are at the same size of the diffraction Airy disk, as shown in Fig.~\ref{fig:reconstructed_6x6_USAF}(a)(b). In this case, an extremely high SNR is needed for reconstruction, as the PSF can be regarded as a frequency filter which invokes severe attenuation on the high-frequency information out of its bandwidth. This requires $ \eta N_B\to 0$ and $G\to \infty$.

\begin{figure}[t]
    \includegraphics[width=\linewidth]{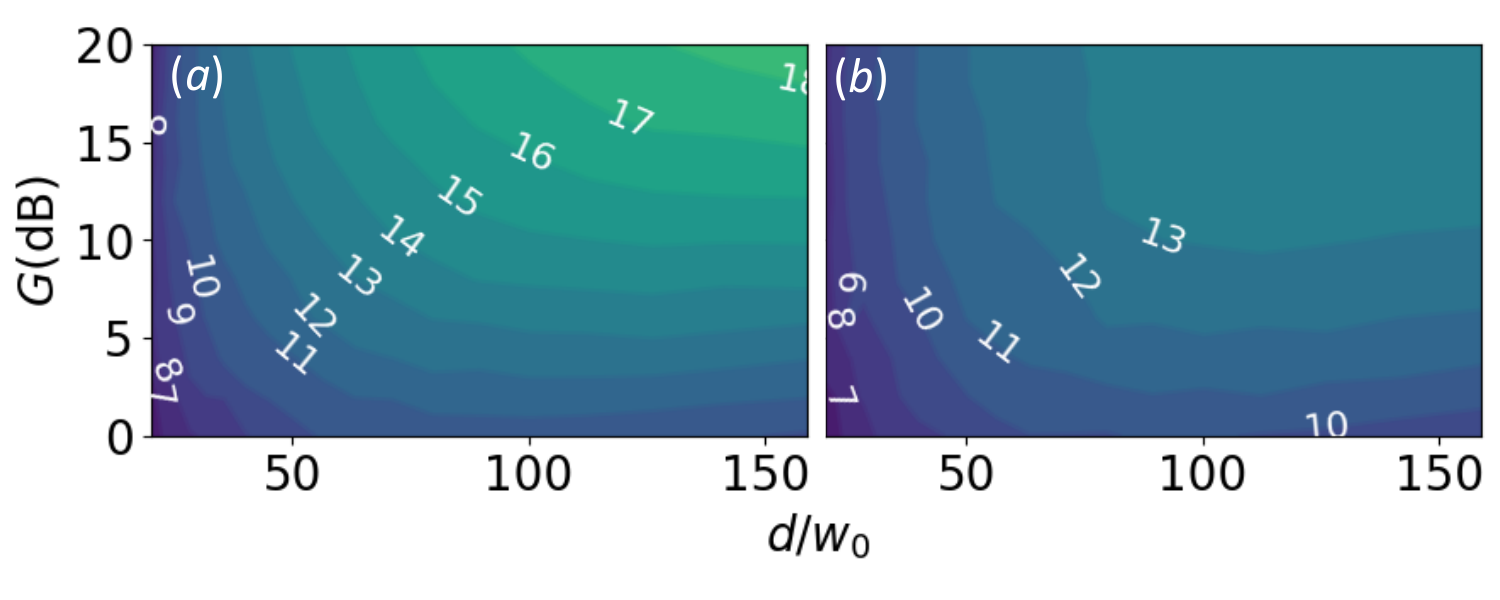}
    \caption{PSNR of the reconstructed images of $n_{P'}=100$. Noise setup: (a) Quantum-limited, $N_B'=0$; (b) thermal, $N_B'=0.1$. 
    }
    \label{fig:PSNR}
\end{figure}

Here we further clarify about diffraction limit. An imaging system has an inevitable diffraction blur due to the finite aperture size. The imaging system is typically linear shift-invariant, thus the blur can be modelled by the PSF which describes the image pattern given a point source, and the image is given by the convolution of the object with the PSF. For an aberration-free imaging system, the resolution is diffraction-limited. The Rayleigh limit gives the resolution limit of a circular aperture. 
It is known that the Rayleigh limit can be surpassed by solving the inverse of the linear imaging system~\cite{frieden1979image,den1997resolution}. For noise-free images, the inverse is perfect. Thus the resolution is limited by solely the image noise, which consists of unknown aberrations and the detection noises~\cite{frieden1967band,frieden1972restoring,frieden1979image}.
To see this, note that the inverse of imaging process is an ill-posed problem~\cite{bertero2021introduction}, in the sense that the imaging process is a multiple-to-one mapping that results in loss of information, due to the finite bandwidth of the imaging system limited by the finite aperture size. As an ill-posed problem, the inverse amplifies any small noises in the image producing a wildly oscillating reconstruction~\cite{bertero2021introduction}. Conventionally, the detection noise is regarded inevitable, e.g. photon counting noise is fundamentally lower bounded by the shot noise limit enforced by quantum mechanics. However, using quantum sources, the noise can be arbitrarily suppressed.
Finally, we point out that our works consider the intermediate-high brightness region, while other works on super-resolution works considers the weak brightness limit~\cite{tsang2016quantum}. There, only the super-resolution beyond Rayleigh limit for two point-source discrimination is considered~\cite{tsang2016quantum}; however it highly relies on the prior knowledge on the object and does not apply to the general imaging scenarios. 



The effect of thermal noise $N_B$ can be seen by comparing the quantum-limited case Fig.~\ref{fig:reconstructed_6x6_USAF}(c) $N_B^\prime=0$ with the thermal case Fig.~\ref{fig:reconstructed_6x6_USAF}(d) $N_B^\prime=0.1$. From Eq.~\eqref{eq:snr} we expect a saturation of $G$ for $G\gtrsim 1/2N_B^\prime$. Note that in subplot (c) the readout variance $\text{var }(\hat X_{P'})=1/4G$, while in subplot (d) $\text{var }(\hat X_{P'})=(1/G+2N_B^\prime)/4$. To achieve the same variance thus match the corresponding contour in subplot (d), in (c) we need $G_c=(1/G_d+2N_B^\prime)^{-1}$. Plugging in $N_B^\prime=0.1$, this means that the contours in (d) are identical with those in (c) for $G_d\ll 1/2N_B^\prime=7$dB, $G_c\ll 1/0.4=4$dB, otherwise we shall see a saturation for $G_d$, specifically contours in (d) resemble those in (c) but stretched significantly for $G_c\to 1/0.2=7$dB. In subplot (d), we verify such saturation for $G$ when $G\gtrsim 4$dB. Also, the contours at $G>4$dB agree with the contours around $4$dB$\le G\le 7$dB in subplot (c) stretched.


\begin{figure}[t]
    \centering
    \includegraphics[width=\linewidth]{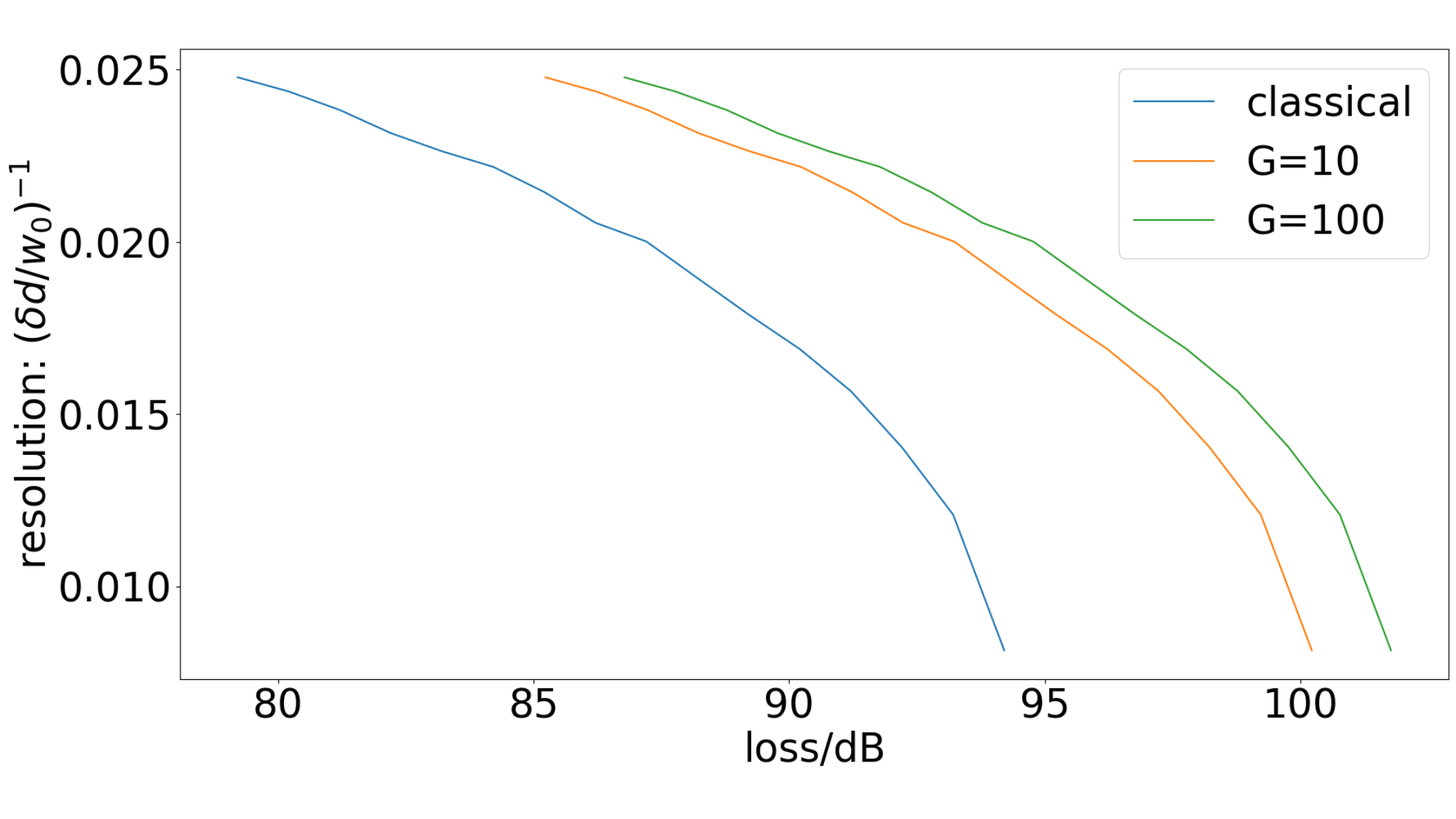}
    \caption{ Resolution versus various sample loss. $n_{P'}=100$. We define the minimum resolvable object size $d_{\rm min}$ as the object size at PSNR$=13$, and we normalize it by the diffraction blur size $w_0=D/2$, $D$ is the physical aperture size. We consider the thermal case with detected dark photon number $N_B'=0.1$. The penetration loss(dB)=$-10\log_{10}\exp{-\alpha z}$, where $z$ is the penetration depth and $\alpha$ is the loss coefficient of the material.}
    \label{fig:resVSloss}
\end{figure}

The size-noise dependence of the PSNR is verified by Fig.~\ref{fig:PSNR}. Here we quantify the reconstructed image quality by the PSNR Eq.~\eqref{eq:psnr}. We can easily verify that the contours are in the similar hyperbola shape as the red curves in Fig.~\ref{fig:reconstructed_6x6_USAF}.

Given an image resolvability criterion that requires ${\rm PSNR}=13$, we plot the resolution in Fig.~\ref{fig:resVSloss}, which is inversely proportional to the minimum resolvable object size $d_{\rm min}$. At the region of higher loss, merely a $10$dB squeezing is sufficient to yield a significant advantage in resolution. Again we identify the saturation of $G$ at $G\gtrsim 1/2N_B'=7$dB, here we see $G=100$dB (orange) makes no significant improvement over $G=10$dB (green).

\section{Conclusion and discussions}
In this work, we propose RF-photonical imaging protocol that benefits from squeezing to enhance the resolution. The protocol is particularly suitable for subterranean imaging scenario, where higher penetration depth requires going to lower frequencies (e.g., ~100kHz) of probing~\cite{korpisalo2014characterization}. A network of antenna connected with fiber-optics is a suitable approach to construct synthetic aperture large enough for the long wavelength (e.g. 10-100 meters). Quantum resource such as squeezing will further enhance the SNR so that the resolution at large penetration depth is acceptable. We have considered imaging where the information about each pixel is extracted to have good resolution, therefore local squeezing at each antenna is needed. In cases where only hypothesis testing is needed, one may be able to have multipartite entanglement to directly enhance the global information extraction for image classification, as studied in Ref.~\cite{zhuang2019physical,xia2021} in the general setting.



\begin{acknowledgements}
The project is supported by Halliburton Technology.
\end{acknowledgements}

\begin{appendix}

\section{Effect of digitalization}
\label{sec:digitalization}

\begin{figure}[t]
    \centering
    \includegraphics[width=\linewidth]{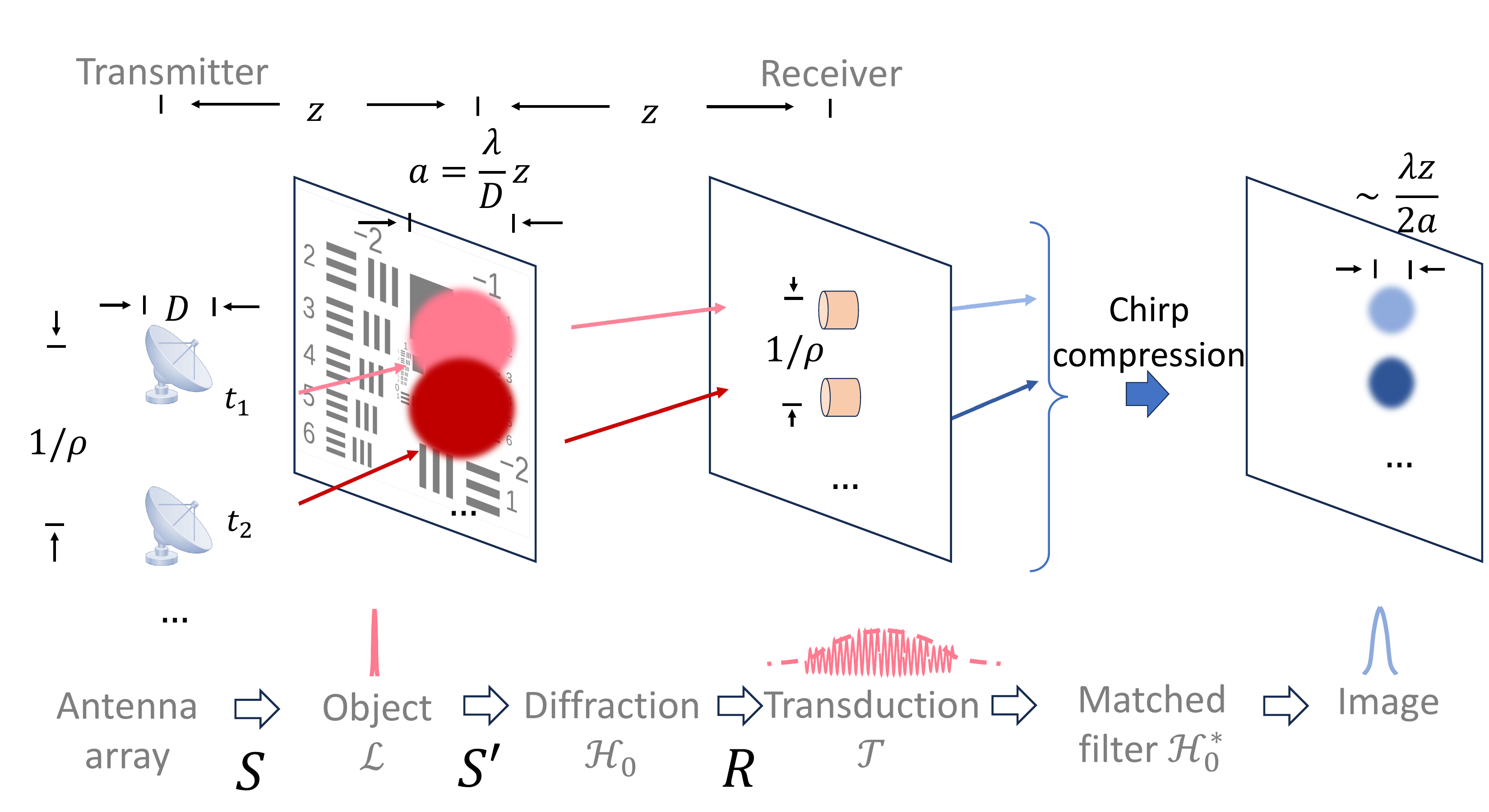}
    \caption{Schematic and parameters of the synthetic aperture radar (SAR). At each time $t$ (or mode index $t$ for MIMO radar), one transmitter shines a beam on the object at distance $z$. Due to the Fresnel diffraction, the corresponding receiver obtains the object pattern mounted on a spatial chirp. Shifting the transmitter and receiver position, the object is imaged to the receiver via a chirp PSF of effective synthetic aperture size (equal to antenna beamwidth) $a\simeq \lambda z/D$. The matched filter fulfills the chirp compression for this PSF, the final compressed PSF has a narrowed blur size $w_0\simeq \lambda z/2a$. A simplified interpretation can be obtained via Fraunhofer approximation $z\to\infty$: the received signal pattern at the receiver (transduction) plane is approximately the Fourier transform of the object, and the matched filter is approximately the inverse Fourier transform. }
    \label{fig:SARschematic}
\end{figure}


We illustrate the detailed schematic of the SAR in Fig.~\ref{fig:SARschematic}. The receiver/transducer plane (\textit{not} the image plane) can be approximately regarded as the Fourier plane of the object (and the image up to a blur), by taking the Fraunhofer approximation at the long propagation distance limit $z\to\infty$. In this case, the image view angle and the image resolution are proportional to the spatial density and the total size of the receiver antenna array respectively.
Let $\rho$ be the spatial density of the antennas (of unit $m^{-1}$), and $L$ be the total size of the antenna array. For a strip-mapping synthetic aperture radar, the receiver and the transmitter have the same spatial arrangement. Below we derive the requirement on the spatial density and total size of practical finite discrete transmitter-receiver arrays.

The object resolution is limited by the worst (minimum) between the size of the beam $a=\lambda z/D$ (synthetic aperture) sent by the transmitter aperture and the total size of the receiver array (Fourier plane size): the minimum resolvable object length $\delta x=\max\{\lambda z/2a,(L/\lambda z)^{-1}\}=\max\{D/2,\lambda z/L\}$.

The view angle only depends on the spatial density of the transmitter-receiver array: $\Delta x/z=(1/\rho/\lambda z)^{-1}/z=\rho\lambda$. 
In the compressed PSF, the distances between the center peak and the closest aliasing peak is $\Delta x$, which gives the maximum view angle $\Delta x/z$. We can verify that a lower sampling rate at the receiver invokes denser aliasing peaks in the PSF (c.f. only one peak in the ideal case), which leads to smaller maximum view angle. 




\end{appendix}


\begin{thebibliography}{42}%
\makeatletter
\providecommand \@ifxundefined [1]{%
 \@ifx{#1\undefined}
}%
\providecommand \@ifnum [1]{%
 \ifnum #1\expandafter \@firstoftwo
 \else \expandafter \@secondoftwo
 \fi
}%
\providecommand \@ifx [1]{%
 \ifx #1\expandafter \@firstoftwo
 \else \expandafter \@secondoftwo
 \fi
}%
\providecommand \natexlab [1]{#1}%
\providecommand \enquote  [1]{``#1''}%
\providecommand \bibnamefont  [1]{#1}%
\providecommand \bibfnamefont [1]{#1}%
\providecommand \citenamefont [1]{#1}%
\providecommand \href@noop [0]{\@secondoftwo}%
\providecommand \href [0]{\begingroup \@sanitize@url \@href}%
\providecommand \@href[1]{\@@startlink{#1}\@@href}%
\providecommand \@@href[1]{\endgroup#1\@@endlink}%
\providecommand \@sanitize@url [0]{\catcode `\\12\catcode `\$12\catcode
  `\&12\catcode `\#12\catcode `\^12\catcode `\_12\catcode `\%12\relax}%
\providecommand \@@startlink[1]{}%
\providecommand \@@endlink[0]{}%
\providecommand \url  [0]{\begingroup\@sanitize@url \@url }%
\providecommand \@url [1]{\endgroup\@href {#1}{\urlprefix }}%
\providecommand \urlprefix  [0]{URL }%
\providecommand \Eprint [0]{\href }%
\providecommand \doibase [0]{https://doi.org/}%
\providecommand \selectlanguage [0]{\@gobble}%
\providecommand \bibinfo  [0]{\@secondoftwo}%
\providecommand \bibfield  [0]{\@secondoftwo}%
\providecommand \translation [1]{[#1]}%
\providecommand \BibitemOpen [0]{}%
\providecommand \bibitemStop [0]{}%
\providecommand \bibitemNoStop [0]{.\EOS\space}%
\providecommand \EOS [0]{\spacefactor3000\relax}%
\providecommand \BibitemShut  [1]{\csname bibitem#1\endcsname}%
\let\auto@bib@innerbib\@empty
\bibitem [{\citenamefont {Giovannetti}\ \emph {et~al.}(2006)\citenamefont
  {Giovannetti}, \citenamefont {Lloyd},\ and\ \citenamefont
  {Maccone}}]{giovannetti2006}%
  \BibitemOpen
  \bibfield  {author} {\bibinfo {author} {\bibfnamefont {V.}~\bibnamefont
  {Giovannetti}}, \bibinfo {author} {\bibfnamefont {S.}~\bibnamefont {Lloyd}},\
  and\ \bibinfo {author} {\bibfnamefont {L.}~\bibnamefont {Maccone}},\ }\href
  {https://doi.org/10.1103/PhysRevLett.96.010401} {\bibfield  {journal}
  {\bibinfo  {journal} {Phys. Rev. Lett.}\ }\textbf {\bibinfo {volume} {96}},\
  \bibinfo {pages} {010401} (\bibinfo {year} {2006})}\BibitemShut {NoStop}%
\bibitem [{\citenamefont {Pirandola}\ \emph {et~al.}(2018)\citenamefont
  {Pirandola}, \citenamefont {Bardhan}, \citenamefont {Gehring}, \citenamefont
  {Weedbrook},\ and\ \citenamefont {Lloyd}}]{pirandola2018advances}%
  \BibitemOpen
  \bibfield  {author} {\bibinfo {author} {\bibfnamefont {S.}~\bibnamefont
  {Pirandola}}, \bibinfo {author} {\bibfnamefont {B.~R.}\ \bibnamefont
  {Bardhan}}, \bibinfo {author} {\bibfnamefont {T.}~\bibnamefont {Gehring}},
  \bibinfo {author} {\bibfnamefont {C.}~\bibnamefont {Weedbrook}},\ and\
  \bibinfo {author} {\bibfnamefont {S.}~\bibnamefont {Lloyd}},\ }\href@noop {}
  {\bibfield  {journal} {\bibinfo  {journal} {Nature Photonics}\ }\textbf
  {\bibinfo {volume} {12}},\ \bibinfo {pages} {724} (\bibinfo {year}
  {2018})}\BibitemShut {NoStop}%
\bibitem [{\citenamefont {Lawrie}\ \emph {et~al.}(2019)\citenamefont {Lawrie},
  \citenamefont {Lett}, \citenamefont {Marino},\ and\ \citenamefont
  {Pooser}}]{lawrie2019quantum}%
  \BibitemOpen
  \bibfield  {author} {\bibinfo {author} {\bibfnamefont {B.~J.}\ \bibnamefont
  {Lawrie}}, \bibinfo {author} {\bibfnamefont {P.~D.}\ \bibnamefont {Lett}},
  \bibinfo {author} {\bibfnamefont {A.~M.}\ \bibnamefont {Marino}},\ and\
  \bibinfo {author} {\bibfnamefont {R.~C.}\ \bibnamefont {Pooser}},\
  }\href@noop {} {\bibfield  {journal} {\bibinfo  {journal} {Acs Photonics}\
  }\textbf {\bibinfo {volume} {6}},\ \bibinfo {pages} {1307} (\bibinfo {year}
  {2019})}\BibitemShut {NoStop}%
\bibitem [{\citenamefont {Zhang}\ and\ \citenamefont
  {Zhuang}(2021)}]{zhang2021dqs}%
  \BibitemOpen
  \bibfield  {author} {\bibinfo {author} {\bibfnamefont {Z.}~\bibnamefont
  {Zhang}}\ and\ \bibinfo {author} {\bibfnamefont {Q.}~\bibnamefont {Zhuang}},\
  }\href {https://doi.org/10.1088/2058-9565/abd4c3} {\bibfield  {journal}
  {\bibinfo  {journal} {Quantum Sci. and Technol.}\ }\textbf {\bibinfo {volume}
  {6}},\ \bibinfo {pages} {043001} (\bibinfo {year} {2021})}\BibitemShut
  {NoStop}%
\bibitem [{\citenamefont {Abadie}\ \emph {et~al.}(2011)\citenamefont {Abadie},
  \citenamefont {Abbott}, \citenamefont {Abbott}, \citenamefont {Abbott},
  \citenamefont {Abernathy}, \citenamefont {Adams}, \citenamefont {Adhikari},
  \citenamefont {Affeldt}, \citenamefont {Allen}, \citenamefont {Allen} \emph
  {et~al.}}]{abadie2011gravitational}%
  \BibitemOpen
  \bibfield  {author} {\bibinfo {author} {\bibfnamefont {J.}~\bibnamefont
  {Abadie}}, \bibinfo {author} {\bibfnamefont {B.~P.}\ \bibnamefont {Abbott}},
  \bibinfo {author} {\bibfnamefont {R.}~\bibnamefont {Abbott}}, \bibinfo
  {author} {\bibfnamefont {T.~D.}\ \bibnamefont {Abbott}}, \bibinfo {author}
  {\bibfnamefont {M.}~\bibnamefont {Abernathy}}, \bibinfo {author}
  {\bibfnamefont {C.}~\bibnamefont {Adams}}, \bibinfo {author} {\bibfnamefont
  {R.}~\bibnamefont {Adhikari}}, \bibinfo {author} {\bibfnamefont
  {C.}~\bibnamefont {Affeldt}}, \bibinfo {author} {\bibfnamefont
  {B.}~\bibnamefont {Allen}}, \bibinfo {author} {\bibfnamefont
  {G.}~\bibnamefont {Allen}}, \emph {et~al.},\ }\href@noop {} {\bibfield
  {journal} {\bibinfo  {journal} {Nat. Phys.}\ }\textbf {\bibinfo {volume}
  {7}},\ \bibinfo {pages} {962} (\bibinfo {year} {2011})}\BibitemShut {NoStop}%
\bibitem [{\citenamefont {Aasi}\ \emph {et~al.}(2013)\citenamefont {Aasi},
  \citenamefont {Abadie}, \citenamefont {Abbott}, \citenamefont {Abbott},
  \citenamefont {Abbott}, \citenamefont {Abernathy}, \citenamefont {Adams},
  \citenamefont {Adams}, \citenamefont {Addesso}, \citenamefont {Adhikari}
  \emph {et~al.}}]{aasi2013enhanced}%
  \BibitemOpen
  \bibfield  {author} {\bibinfo {author} {\bibfnamefont {J.}~\bibnamefont
  {Aasi}}, \bibinfo {author} {\bibfnamefont {J.}~\bibnamefont {Abadie}},
  \bibinfo {author} {\bibfnamefont {B.}~\bibnamefont {Abbott}}, \bibinfo
  {author} {\bibfnamefont {R.}~\bibnamefont {Abbott}}, \bibinfo {author}
  {\bibfnamefont {T.}~\bibnamefont {Abbott}}, \bibinfo {author} {\bibfnamefont
  {M.}~\bibnamefont {Abernathy}}, \bibinfo {author} {\bibfnamefont
  {C.}~\bibnamefont {Adams}}, \bibinfo {author} {\bibfnamefont
  {T.}~\bibnamefont {Adams}}, \bibinfo {author} {\bibfnamefont
  {P.}~\bibnamefont {Addesso}}, \bibinfo {author} {\bibfnamefont
  {R.}~\bibnamefont {Adhikari}}, \emph {et~al.},\ }\href@noop {} {\bibfield
  {journal} {\bibinfo  {journal} {Nat. Photonics}\ }\textbf {\bibinfo {volume}
  {7}},\ \bibinfo {pages} {613} (\bibinfo {year} {2013})}\BibitemShut {NoStop}%
\bibitem [{\citenamefont {Tse}\ \emph {et~al.}(2019)\citenamefont {Tse},
  \citenamefont {Yu}, \citenamefont {Kijbunchoo}, \citenamefont
  {Fernandez-Galiana}, \citenamefont {Dupej}, \citenamefont {Barsotti},
  \citenamefont {Blair}, \citenamefont {Brown}, \citenamefont {Dwyer},
  \citenamefont {Effler} \emph {et~al.}}]{tse2019quantum}%
  \BibitemOpen
  \bibfield  {author} {\bibinfo {author} {\bibfnamefont {M.}~\bibnamefont
  {Tse}}, \bibinfo {author} {\bibfnamefont {H.}~\bibnamefont {Yu}}, \bibinfo
  {author} {\bibfnamefont {N.}~\bibnamefont {Kijbunchoo}}, \bibinfo {author}
  {\bibfnamefont {A.}~\bibnamefont {Fernandez-Galiana}}, \bibinfo {author}
  {\bibfnamefont {P.}~\bibnamefont {Dupej}}, \bibinfo {author} {\bibfnamefont
  {L.}~\bibnamefont {Barsotti}}, \bibinfo {author} {\bibfnamefont
  {C.}~\bibnamefont {Blair}}, \bibinfo {author} {\bibfnamefont
  {D.}~\bibnamefont {Brown}}, \bibinfo {author} {\bibfnamefont
  {S.}~\bibnamefont {Dwyer}}, \bibinfo {author} {\bibfnamefont
  {A.}~\bibnamefont {Effler}}, \emph {et~al.},\ }\href@noop {} {\bibfield
  {journal} {\bibinfo  {journal} {Phys. Rev. Lett.}\ }\textbf {\bibinfo
  {volume} {123}},\ \bibinfo {pages} {231107} (\bibinfo {year}
  {2019})}\BibitemShut {NoStop}%
\bibitem [{\citenamefont {Backes}\ \emph {et~al.}(2021)\citenamefont {Backes},
  \citenamefont {Palken}, \citenamefont {Al~Kenany}, \citenamefont {Brubaker},
  \citenamefont {Cahn}, \citenamefont {Droster}, \citenamefont {Hilton},
  \citenamefont {Ghosh}, \citenamefont {Jackson}, \citenamefont {Lamoreaux}
  \emph {et~al.}}]{backes2021}%
  \BibitemOpen
  \bibfield  {author} {\bibinfo {author} {\bibfnamefont {K.}~\bibnamefont
  {Backes}}, \bibinfo {author} {\bibfnamefont {D.}~\bibnamefont {Palken}},
  \bibinfo {author} {\bibfnamefont {S.}~\bibnamefont {Al~Kenany}}, \bibinfo
  {author} {\bibfnamefont {B.}~\bibnamefont {Brubaker}}, \bibinfo {author}
  {\bibfnamefont {S.}~\bibnamefont {Cahn}}, \bibinfo {author} {\bibfnamefont
  {A.}~\bibnamefont {Droster}}, \bibinfo {author} {\bibfnamefont {G.~C.}\
  \bibnamefont {Hilton}}, \bibinfo {author} {\bibfnamefont {S.}~\bibnamefont
  {Ghosh}}, \bibinfo {author} {\bibfnamefont {H.}~\bibnamefont {Jackson}},
  \bibinfo {author} {\bibfnamefont {S.}~\bibnamefont {Lamoreaux}}, \emph
  {et~al.},\ }\href {https://doi.org/10.1038/s41586-021-03226-7} {\bibfield
  {journal} {\bibinfo  {journal} {Nature}\ }\textbf {\bibinfo {volume} {590}},\
  \bibinfo {pages} {238} (\bibinfo {year} {2021})}\BibitemShut {NoStop}%
\bibitem [{\citenamefont {Giovannetti}\ \emph {et~al.}(2001)\citenamefont
  {Giovannetti}, \citenamefont {Lloyd},\ and\ \citenamefont
  {Maccone}}]{giovannetti2001quantum}%
  \BibitemOpen
  \bibfield  {author} {\bibinfo {author} {\bibfnamefont {V.}~\bibnamefont
  {Giovannetti}}, \bibinfo {author} {\bibfnamefont {S.}~\bibnamefont {Lloyd}},\
  and\ \bibinfo {author} {\bibfnamefont {L.}~\bibnamefont {Maccone}},\
  }\href@noop {} {\bibfield  {journal} {\bibinfo  {journal} {Nature}\ }\textbf
  {\bibinfo {volume} {412}},\ \bibinfo {pages} {417} (\bibinfo {year}
  {2001})}\BibitemShut {NoStop}%
\bibitem [{\citenamefont {Zhuang}\ \emph {et~al.}(2017)\citenamefont {Zhuang},
  \citenamefont {Zhang},\ and\ \citenamefont
  {Shapiro}}]{zhuang2017entanglement}%
  \BibitemOpen
  \bibfield  {author} {\bibinfo {author} {\bibfnamefont {Q.}~\bibnamefont
  {Zhuang}}, \bibinfo {author} {\bibfnamefont {Z.}~\bibnamefont {Zhang}},\ and\
  \bibinfo {author} {\bibfnamefont {J.~H.}\ \bibnamefont {Shapiro}},\
  }\href@noop {} {\bibfield  {journal} {\bibinfo  {journal} {Phys. Rev. A}\
  }\textbf {\bibinfo {volume} {96}},\ \bibinfo {pages} {040304} (\bibinfo
  {year} {2017})}\BibitemShut {NoStop}%
\bibitem [{\citenamefont {Huang}\ \emph {et~al.}(2021)\citenamefont {Huang},
  \citenamefont {Lupo},\ and\ \citenamefont {Kok}}]{huang2021quantum}%
  \BibitemOpen
  \bibfield  {author} {\bibinfo {author} {\bibfnamefont {Z.}~\bibnamefont
  {Huang}}, \bibinfo {author} {\bibfnamefont {C.}~\bibnamefont {Lupo}},\ and\
  \bibinfo {author} {\bibfnamefont {P.}~\bibnamefont {Kok}},\ }\href@noop {}
  {\bibfield  {journal} {\bibinfo  {journal} {PRX Quantum}\ }\textbf {\bibinfo
  {volume} {2}},\ \bibinfo {pages} {030303} (\bibinfo {year}
  {2021})}\BibitemShut {NoStop}%
\bibitem [{\citenamefont {Reichert}\ \emph {et~al.}(2022)\citenamefont
  {Reichert}, \citenamefont {Di~Candia}, \citenamefont {Win},\ and\
  \citenamefont {Sanz}}]{reichert2022quantum}%
  \BibitemOpen
  \bibfield  {author} {\bibinfo {author} {\bibfnamefont {M.}~\bibnamefont
  {Reichert}}, \bibinfo {author} {\bibfnamefont {R.}~\bibnamefont {Di~Candia}},
  \bibinfo {author} {\bibfnamefont {M.~Z.}\ \bibnamefont {Win}},\ and\ \bibinfo
  {author} {\bibfnamefont {M.}~\bibnamefont {Sanz}},\ }\href@noop {} {\bibfield
   {journal} {\bibinfo  {journal} {npj Quantum Information}\ }\textbf {\bibinfo
  {volume} {8}},\ \bibinfo {pages} {147} (\bibinfo {year} {2022})}\BibitemShut
  {NoStop}%
\bibitem [{\citenamefont {Reichert}\ \emph {et~al.}(2023)\citenamefont
  {Reichert}, \citenamefont {Zhuang},\ and\ \citenamefont
  {Sanz}}]{reichert2023heisenberg}%
  \BibitemOpen
  \bibfield  {author} {\bibinfo {author} {\bibfnamefont {M.}~\bibnamefont
  {Reichert}}, \bibinfo {author} {\bibfnamefont {Q.}~\bibnamefont {Zhuang}},\
  and\ \bibinfo {author} {\bibfnamefont {M.}~\bibnamefont {Sanz}},\ }\href@noop
  {} {\bibfield  {journal} {\bibinfo  {journal} {arXiv:2311.14546}\ } (\bibinfo
  {year} {2023})}\BibitemShut {NoStop}%
\bibitem [{\citenamefont {Tan}\ \emph {et~al.}(2008)\citenamefont {Tan},
  \citenamefont {Erkmen}, \citenamefont {Giovannetti}, \citenamefont {Guha},
  \citenamefont {Lloyd}, \citenamefont {Maccone}, \citenamefont {Pirandola},\
  and\ \citenamefont {Shapiro}}]{tan2008quantum}%
  \BibitemOpen
  \bibfield  {author} {\bibinfo {author} {\bibfnamefont {S.-H.}\ \bibnamefont
  {Tan}}, \bibinfo {author} {\bibfnamefont {B.~I.}\ \bibnamefont {Erkmen}},
  \bibinfo {author} {\bibfnamefont {V.}~\bibnamefont {Giovannetti}}, \bibinfo
  {author} {\bibfnamefont {S.}~\bibnamefont {Guha}}, \bibinfo {author}
  {\bibfnamefont {S.}~\bibnamefont {Lloyd}}, \bibinfo {author} {\bibfnamefont
  {L.}~\bibnamefont {Maccone}}, \bibinfo {author} {\bibfnamefont
  {S.}~\bibnamefont {Pirandola}},\ and\ \bibinfo {author} {\bibfnamefont
  {J.~H.}\ \bibnamefont {Shapiro}},\ }\href@noop {} {\bibfield  {journal}
  {\bibinfo  {journal} {Phys. Rev. Lett.}\ }\textbf {\bibinfo {volume} {101}},\
  \bibinfo {pages} {253601} (\bibinfo {year} {2008})}\BibitemShut {NoStop}%
\bibitem [{\citenamefont {Zhang}\ \emph {et~al.}(2015)\citenamefont {Zhang},
  \citenamefont {Mouradian}, \citenamefont {Wong},\ and\ \citenamefont
  {Shapiro}}]{zhang2015}%
  \BibitemOpen
  \bibfield  {author} {\bibinfo {author} {\bibfnamefont {Z.}~\bibnamefont
  {Zhang}}, \bibinfo {author} {\bibfnamefont {S.}~\bibnamefont {Mouradian}},
  \bibinfo {author} {\bibfnamefont {F.~N.~C.}\ \bibnamefont {Wong}},\ and\
  \bibinfo {author} {\bibfnamefont {J.~H.}\ \bibnamefont {Shapiro}},\ }\href
  {https://doi.org/10.1103/PhysRevLett.114.110506} {\bibfield  {journal}
  {\bibinfo  {journal} {Phys. Rev. Lett.}\ }\textbf {\bibinfo {volume} {114}},\
  \bibinfo {pages} {110506} (\bibinfo {year} {2015})}\BibitemShut {NoStop}%
\bibitem [{\citenamefont {Assouly}\ \emph {et~al.}(2023)\citenamefont
  {Assouly}, \citenamefont {Dassonneville}, \citenamefont {Peronnin},
  \citenamefont {Bienfait},\ and\ \citenamefont {Huard}}]{assouly2023quantum}%
  \BibitemOpen
  \bibfield  {author} {\bibinfo {author} {\bibfnamefont {R.}~\bibnamefont
  {Assouly}}, \bibinfo {author} {\bibfnamefont {R.}~\bibnamefont
  {Dassonneville}}, \bibinfo {author} {\bibfnamefont {T.}~\bibnamefont
  {Peronnin}}, \bibinfo {author} {\bibfnamefont {A.}~\bibnamefont {Bienfait}},\
  and\ \bibinfo {author} {\bibfnamefont {B.}~\bibnamefont {Huard}},\
  }\href@noop {} {\bibfield  {journal} {\bibinfo  {journal} {Nature Physics}\
  ,\ \bibinfo {pages} {1}} (\bibinfo {year} {2023})}\BibitemShut {NoStop}%
\bibitem [{\citenamefont {Zhuang}\ and\ \citenamefont
  {Shapiro}(2022)}]{zhuang2022}%
  \BibitemOpen
  \bibfield  {author} {\bibinfo {author} {\bibfnamefont {Q.}~\bibnamefont
  {Zhuang}}\ and\ \bibinfo {author} {\bibfnamefont {J.~H.}\ \bibnamefont
  {Shapiro}},\ }\href {https://doi.org/10.1103/PhysRevLett.128.010501}
  {\bibfield  {journal} {\bibinfo  {journal} {Phys. Rev. Lett.}\ }\textbf
  {\bibinfo {volume} {128}},\ \bibinfo {pages} {010501} (\bibinfo {year}
  {2022})}\BibitemShut {NoStop}%
\bibitem [{\citenamefont {Shapiro}(2020)}]{shapiro2020quantum}%
  \BibitemOpen
  \bibfield  {author} {\bibinfo {author} {\bibfnamefont {J.~H.}\ \bibnamefont
  {Shapiro}},\ }\href@noop {} {\bibfield  {journal} {\bibinfo  {journal} {IEEE
  Aerospace and Electronic Systems Magazine}\ }\textbf {\bibinfo {volume}
  {35}},\ \bibinfo {pages} {8} (\bibinfo {year} {2020})}\BibitemShut {NoStop}%
\bibitem [{\citenamefont {Xia}\ \emph {et~al.}(2020)\citenamefont {Xia},
  \citenamefont {Li}, \citenamefont {Clark}, \citenamefont {Hart},
  \citenamefont {Zhuang},\ and\ \citenamefont {Zhang}}]{xia2020demonstration}%
  \BibitemOpen
  \bibfield  {author} {\bibinfo {author} {\bibfnamefont {Y.}~\bibnamefont
  {Xia}}, \bibinfo {author} {\bibfnamefont {W.}~\bibnamefont {Li}}, \bibinfo
  {author} {\bibfnamefont {W.}~\bibnamefont {Clark}}, \bibinfo {author}
  {\bibfnamefont {D.}~\bibnamefont {Hart}}, \bibinfo {author} {\bibfnamefont
  {Q.}~\bibnamefont {Zhuang}},\ and\ \bibinfo {author} {\bibfnamefont
  {Z.}~\bibnamefont {Zhang}},\ }\href@noop {} {\bibfield  {journal} {\bibinfo
  {journal} {Phys. Rev. Lett.}\ }\textbf {\bibinfo {volume} {124}},\ \bibinfo
  {pages} {150502} (\bibinfo {year} {2020})}\BibitemShut {NoStop}%
\bibitem [{\citenamefont {Korpisalo}\ \emph {et~al.}(2014)\citenamefont
  {Korpisalo} \emph {et~al.}}]{korpisalo2014characterization}%
  \BibitemOpen
  \bibfield  {author} {\bibinfo {author} {\bibfnamefont {A.}~\bibnamefont
  {Korpisalo}} \emph {et~al.},\ }\href@noop {} {\bibfield  {journal} {\bibinfo
  {journal} {International Journal of Geophysics}\ }\textbf {\bibinfo {volume}
  {2014}} (\bibinfo {year} {2014})}\BibitemShut {NoStop}%
\bibitem [{\citenamefont {Franceschetti}\ and\ \citenamefont
  {Lanari}(2018)}]{franceschetti2018synthetic}%
  \BibitemOpen
  \bibfield  {author} {\bibinfo {author} {\bibfnamefont {G.}~\bibnamefont
  {Franceschetti}}\ and\ \bibinfo {author} {\bibfnamefont {R.}~\bibnamefont
  {Lanari}},\ }\href@noop {} {\emph {\bibinfo {title} {Synthetic aperture radar
  processing}}}\ (\bibinfo  {publisher} {CRC press},\ \bibinfo {year}
  {2018})\BibitemShut {NoStop}%
\bibitem [{\citenamefont {Chapin}\ \emph {et~al.}(2012)\citenamefont {Chapin},
  \citenamefont {Chau}, \citenamefont {Chen}, \citenamefont {Heavey},
  \citenamefont {Hensley}, \citenamefont {Lou}, \citenamefont {Machuzak},\ and\
  \citenamefont {Moghaddam}}]{chapin2012airmoss}%
  \BibitemOpen
  \bibfield  {author} {\bibinfo {author} {\bibfnamefont {E.}~\bibnamefont
  {Chapin}}, \bibinfo {author} {\bibfnamefont {A.}~\bibnamefont {Chau}},
  \bibinfo {author} {\bibfnamefont {J.}~\bibnamefont {Chen}}, \bibinfo {author}
  {\bibfnamefont {B.}~\bibnamefont {Heavey}}, \bibinfo {author} {\bibfnamefont
  {S.}~\bibnamefont {Hensley}}, \bibinfo {author} {\bibfnamefont
  {Y.}~\bibnamefont {Lou}}, \bibinfo {author} {\bibfnamefont {R.}~\bibnamefont
  {Machuzak}},\ and\ \bibinfo {author} {\bibfnamefont {M.}~\bibnamefont
  {Moghaddam}},\ }in\ \href@noop {} {\emph {\bibinfo {booktitle} {2012 IEEE
  radar conference}}}\ (\bibinfo {organization} {IEEE},\ \bibinfo {year}
  {2012})\ pp.\ \bibinfo {pages} {0693--0698}\BibitemShut {NoStop}%
\bibitem [{\citenamefont {Krieger}\ \emph {et~al.}(2007)\citenamefont
  {Krieger}, \citenamefont {Gebert},\ and\ \citenamefont
  {Moreira}}]{krieger2007multidimensional}%
  \BibitemOpen
  \bibfield  {author} {\bibinfo {author} {\bibfnamefont {G.}~\bibnamefont
  {Krieger}}, \bibinfo {author} {\bibfnamefont {N.}~\bibnamefont {Gebert}},\
  and\ \bibinfo {author} {\bibfnamefont {A.}~\bibnamefont {Moreira}},\
  }\href@noop {} {\bibfield  {journal} {\bibinfo  {journal} {IEEE Transactions
  on Geoscience and Remote Sensing}\ }\textbf {\bibinfo {volume} {46}},\
  \bibinfo {pages} {31} (\bibinfo {year} {2007})}\BibitemShut {NoStop}%
\bibitem [{\citenamefont {Ghelfi}\ \emph {et~al.}(2014)\citenamefont {Ghelfi},
  \citenamefont {Laghezza}, \citenamefont {Scotti}, \citenamefont {Serafino},
  \citenamefont {Capria}, \citenamefont {Pinna}, \citenamefont {Onori},
  \citenamefont {Porzi}, \citenamefont {Scaffardi}, \citenamefont {Malacarne}
  \emph {et~al.}}]{ghelfi2014fully}%
  \BibitemOpen
  \bibfield  {author} {\bibinfo {author} {\bibfnamefont {P.}~\bibnamefont
  {Ghelfi}}, \bibinfo {author} {\bibfnamefont {F.}~\bibnamefont {Laghezza}},
  \bibinfo {author} {\bibfnamefont {F.}~\bibnamefont {Scotti}}, \bibinfo
  {author} {\bibfnamefont {G.}~\bibnamefont {Serafino}}, \bibinfo {author}
  {\bibfnamefont {A.}~\bibnamefont {Capria}}, \bibinfo {author} {\bibfnamefont
  {S.}~\bibnamefont {Pinna}}, \bibinfo {author} {\bibfnamefont
  {D.}~\bibnamefont {Onori}}, \bibinfo {author} {\bibfnamefont
  {C.}~\bibnamefont {Porzi}}, \bibinfo {author} {\bibfnamefont
  {M.}~\bibnamefont {Scaffardi}}, \bibinfo {author} {\bibfnamefont
  {A.}~\bibnamefont {Malacarne}}, \emph {et~al.},\ }\href@noop {} {\bibfield
  {journal} {\bibinfo  {journal} {Nature}\ }\textbf {\bibinfo {volume} {507}},\
  \bibinfo {pages} {341} (\bibinfo {year} {2014})}\BibitemShut {NoStop}%
\bibitem [{\citenamefont {Zhang}\ \emph {et~al.}(2017)\citenamefont {Zhang},
  \citenamefont {Guo}, \citenamefont {Wang}, \citenamefont {Zhou},
  \citenamefont {Zhang}, \citenamefont {Sun},\ and\ \citenamefont
  {Pan}}]{zhang2017photonics}%
  \BibitemOpen
  \bibfield  {author} {\bibinfo {author} {\bibfnamefont {F.}~\bibnamefont
  {Zhang}}, \bibinfo {author} {\bibfnamefont {Q.}~\bibnamefont {Guo}}, \bibinfo
  {author} {\bibfnamefont {Z.}~\bibnamefont {Wang}}, \bibinfo {author}
  {\bibfnamefont {P.}~\bibnamefont {Zhou}}, \bibinfo {author} {\bibfnamefont
  {G.}~\bibnamefont {Zhang}}, \bibinfo {author} {\bibfnamefont
  {J.}~\bibnamefont {Sun}},\ and\ \bibinfo {author} {\bibfnamefont
  {S.}~\bibnamefont {Pan}},\ }\href@noop {} {\bibfield  {journal} {\bibinfo
  {journal} {Optics Express}\ }\textbf {\bibinfo {volume} {25}},\ \bibinfo
  {pages} {16274} (\bibinfo {year} {2017})}\BibitemShut {NoStop}%
\bibitem [{\citenamefont {Dong}\ \emph {et~al.}(2020)\citenamefont {Dong},
  \citenamefont {Zhang}, \citenamefont {Jiao}, \citenamefont {Sun},\ and\
  \citenamefont {Li}}]{dong2020microwave}%
  \BibitemOpen
  \bibfield  {author} {\bibinfo {author} {\bibfnamefont {J.}~\bibnamefont
  {Dong}}, \bibinfo {author} {\bibfnamefont {F.}~\bibnamefont {Zhang}},
  \bibinfo {author} {\bibfnamefont {Z.}~\bibnamefont {Jiao}}, \bibinfo {author}
  {\bibfnamefont {Q.}~\bibnamefont {Sun}},\ and\ \bibinfo {author}
  {\bibfnamefont {W.}~\bibnamefont {Li}},\ }\href@noop {} {\bibfield  {journal}
  {\bibinfo  {journal} {Optics Express}\ }\textbf {\bibinfo {volume} {28}},\
  \bibinfo {pages} {19113} (\bibinfo {year} {2020})}\BibitemShut {NoStop}%
\bibitem [{\citenamefont {Frieden}(1979)}]{frieden1979image}%
  \BibitemOpen
  \bibfield  {author} {\bibinfo {author} {\bibfnamefont {B.}~\bibnamefont
  {Frieden}},\ }\href@noop {} {\bibfield  {journal} {\bibinfo  {journal}
  {Picture Processing and Digital Filtering}\ }\textbf {\bibinfo {volume}
  {6}},\ \bibinfo {pages} {177} (\bibinfo {year} {1979})}\BibitemShut {NoStop}%
\bibitem [{\citenamefont {Den~Dekker}\ and\ \citenamefont {Van~den
  Bos}(1997)}]{den1997resolution}%
  \BibitemOpen
  \bibfield  {author} {\bibinfo {author} {\bibfnamefont {A.~J.}\ \bibnamefont
  {Den~Dekker}}\ and\ \bibinfo {author} {\bibfnamefont {A.}~\bibnamefont
  {Van~den Bos}},\ }\href@noop {} {\bibfield  {journal} {\bibinfo  {journal}
  {JOSA A}\ }\textbf {\bibinfo {volume} {14}},\ \bibinfo {pages} {547}
  (\bibinfo {year} {1997})}\BibitemShut {NoStop}%
\bibitem [{\citenamefont {Frieden}(1967)}]{frieden1967band}%
  \BibitemOpen
  \bibfield  {author} {\bibinfo {author} {\bibfnamefont {B.~R.}\ \bibnamefont
  {Frieden}},\ }\href@noop {} {\bibfield  {journal} {\bibinfo  {journal}
  {JOSA}\ }\textbf {\bibinfo {volume} {57}},\ \bibinfo {pages} {1013} (\bibinfo
  {year} {1967})}\BibitemShut {NoStop}%
\bibitem [{\citenamefont {Frieden}(1972)}]{frieden1972restoring}%
  \BibitemOpen
  \bibfield  {author} {\bibinfo {author} {\bibfnamefont {B.~R.}\ \bibnamefont
  {Frieden}},\ }\href@noop {} {\bibfield  {journal} {\bibinfo  {journal}
  {JOSA}\ }\textbf {\bibinfo {volume} {62}},\ \bibinfo {pages} {511} (\bibinfo
  {year} {1972})}\BibitemShut {NoStop}%
\bibitem [{\citenamefont {Salomon}(2002)}]{salomon2002data}%
  \BibitemOpen
  \bibfield  {author} {\bibinfo {author} {\bibfnamefont {D.}~\bibnamefont
  {Salomon}},\ }\href@noop {} {\emph {\bibinfo {title} {Data compression}}}\
  (\bibinfo  {publisher} {Springer},\ \bibinfo {year} {2002})\BibitemShut
  {NoStop}%
\bibitem [{\citenamefont {Munson}\ and\ \citenamefont
  {Visentin}(1989)}]{munson1989signal}%
  \BibitemOpen
  \bibfield  {author} {\bibinfo {author} {\bibfnamefont {D.~C.}\ \bibnamefont
  {Munson}}\ and\ \bibinfo {author} {\bibfnamefont {R.~L.}\ \bibnamefont
  {Visentin}},\ }\href@noop {} {\bibfield  {journal} {\bibinfo  {journal} {IEEE
  Transactions on Acoustics, Speech, and Signal Processing}\ }\textbf {\bibinfo
  {volume} {37}},\ \bibinfo {pages} {2131} (\bibinfo {year}
  {1989})}\BibitemShut {NoStop}%
\bibitem [{\citenamefont {Staelin}(2011)}]{staelin2011electromagnetics}%
  \BibitemOpen
  \bibfield  {author} {\bibinfo {author} {\bibfnamefont {D.~H.}\ \bibnamefont
  {Staelin}},\ }\href@noop {} {\emph {\bibinfo {title} {Electromagnetics and
  applications}}}\ (\bibinfo  {publisher} {Massachusetts Institute of
  Technology Cambridge, MA, USA},\ \bibinfo {year} {2011})\BibitemShut
  {NoStop}%
\bibitem [{\citenamefont {Hao}\ \emph {et~al.}(2021)\citenamefont {Hao},
  \citenamefont {Shi}, \citenamefont {Li}, \citenamefont {Shapiro},
  \citenamefont {Zhuang},\ and\ \citenamefont {Zhang}}]{hao2021entanglement}%
  \BibitemOpen
  \bibfield  {author} {\bibinfo {author} {\bibfnamefont {S.}~\bibnamefont
  {Hao}}, \bibinfo {author} {\bibfnamefont {H.}~\bibnamefont {Shi}}, \bibinfo
  {author} {\bibfnamefont {W.}~\bibnamefont {Li}}, \bibinfo {author}
  {\bibfnamefont {J.~H.}\ \bibnamefont {Shapiro}}, \bibinfo {author}
  {\bibfnamefont {Q.}~\bibnamefont {Zhuang}},\ and\ \bibinfo {author}
  {\bibfnamefont {Z.}~\bibnamefont {Zhang}},\ }\href@noop {} {\bibfield
  {journal} {\bibinfo  {journal} {Physical Review Letters}\ }\textbf {\bibinfo
  {volume} {126}},\ \bibinfo {pages} {250501} (\bibinfo {year}
  {2021})}\BibitemShut {NoStop}%
\bibitem [{\citenamefont {Hao}\ \emph {et~al.}(2022)\citenamefont {Hao},
  \citenamefont {Shi}, \citenamefont {Gagatsos}, \citenamefont {Mishra},
  \citenamefont {Bash}, \citenamefont {Djordjevic}, \citenamefont {Guha},
  \citenamefont {Zhuang},\ and\ \citenamefont {Zhang}}]{hao2022demonstration}%
  \BibitemOpen
  \bibfield  {author} {\bibinfo {author} {\bibfnamefont {S.}~\bibnamefont
  {Hao}}, \bibinfo {author} {\bibfnamefont {H.}~\bibnamefont {Shi}}, \bibinfo
  {author} {\bibfnamefont {C.~N.}\ \bibnamefont {Gagatsos}}, \bibinfo {author}
  {\bibfnamefont {M.}~\bibnamefont {Mishra}}, \bibinfo {author} {\bibfnamefont
  {B.}~\bibnamefont {Bash}}, \bibinfo {author} {\bibfnamefont {I.}~\bibnamefont
  {Djordjevic}}, \bibinfo {author} {\bibfnamefont {S.}~\bibnamefont {Guha}},
  \bibinfo {author} {\bibfnamefont {Q.}~\bibnamefont {Zhuang}},\ and\ \bibinfo
  {author} {\bibfnamefont {Z.}~\bibnamefont {Zhang}},\ }\href@noop {}
  {\bibfield  {journal} {\bibinfo  {journal} {Physical Review Letters}\
  }\textbf {\bibinfo {volume} {129}},\ \bibinfo {pages} {010501} (\bibinfo
  {year} {2022})}\BibitemShut {NoStop}%
\bibitem [{\citenamefont {Vahlbruch}\ \emph {et~al.}(2016)\citenamefont
  {Vahlbruch}, \citenamefont {Mehmet}, \citenamefont {Danzmann},\ and\
  \citenamefont {Schnabel}}]{vahlbruch2016}%
  \BibitemOpen
  \bibfield  {author} {\bibinfo {author} {\bibfnamefont {H.}~\bibnamefont
  {Vahlbruch}}, \bibinfo {author} {\bibfnamefont {M.}~\bibnamefont {Mehmet}},
  \bibinfo {author} {\bibfnamefont {K.}~\bibnamefont {Danzmann}},\ and\
  \bibinfo {author} {\bibfnamefont {R.}~\bibnamefont {Schnabel}},\ }\href
  {https://doi.org/10.1103/PhysRevLett.117.110801} {\bibfield  {journal}
  {\bibinfo  {journal} {Phys. Rev. Lett.}\ }\textbf {\bibinfo {volume} {117}},\
  \bibinfo {pages} {110801} (\bibinfo {year} {2016})}\BibitemShut {NoStop}%
\bibitem [{\citenamefont {Weedbrook}\ \emph {et~al.}(2012)\citenamefont
  {Weedbrook}, \citenamefont {Pirandola}, \citenamefont
  {Garc{\'\i}a-Patr{\'o}n}, \citenamefont {Cerf}, \citenamefont {Ralph},
  \citenamefont {Shapiro},\ and\ \citenamefont
  {Lloyd}}]{weedbrook2012gaussian}%
  \BibitemOpen
  \bibfield  {author} {\bibinfo {author} {\bibfnamefont {C.}~\bibnamefont
  {Weedbrook}}, \bibinfo {author} {\bibfnamefont {S.}~\bibnamefont
  {Pirandola}}, \bibinfo {author} {\bibfnamefont {R.}~\bibnamefont
  {Garc{\'\i}a-Patr{\'o}n}}, \bibinfo {author} {\bibfnamefont {N.~J.}\
  \bibnamefont {Cerf}}, \bibinfo {author} {\bibfnamefont {T.~C.}\ \bibnamefont
  {Ralph}}, \bibinfo {author} {\bibfnamefont {J.~H.}\ \bibnamefont {Shapiro}},\
  and\ \bibinfo {author} {\bibfnamefont {S.}~\bibnamefont {Lloyd}},\
  }\href@noop {} {\bibfield  {journal} {\bibinfo  {journal} {Reviews of Modern
  Physics}\ }\textbf {\bibinfo {volume} {84}},\ \bibinfo {pages} {621}
  (\bibinfo {year} {2012})}\BibitemShut {NoStop}%
\bibitem [{\citenamefont {Holzgrafe}\ \emph {et~al.}(2020)\citenamefont
  {Holzgrafe}, \citenamefont {Sinclair}, \citenamefont {Zhu}, \citenamefont
  {Shams-Ansari}, \citenamefont {Colangelo}, \citenamefont {Hu}, \citenamefont
  {Zhang}, \citenamefont {Berggren},\ and\ \citenamefont
  {Lon{\v{c}}ar}}]{holzgrafe2020cavity}%
  \BibitemOpen
  \bibfield  {author} {\bibinfo {author} {\bibfnamefont {J.}~\bibnamefont
  {Holzgrafe}}, \bibinfo {author} {\bibfnamefont {N.}~\bibnamefont {Sinclair}},
  \bibinfo {author} {\bibfnamefont {D.}~\bibnamefont {Zhu}}, \bibinfo {author}
  {\bibfnamefont {A.}~\bibnamefont {Shams-Ansari}}, \bibinfo {author}
  {\bibfnamefont {M.}~\bibnamefont {Colangelo}}, \bibinfo {author}
  {\bibfnamefont {Y.}~\bibnamefont {Hu}}, \bibinfo {author} {\bibfnamefont
  {M.}~\bibnamefont {Zhang}}, \bibinfo {author} {\bibfnamefont {K.~K.}\
  \bibnamefont {Berggren}},\ and\ \bibinfo {author} {\bibfnamefont
  {M.}~\bibnamefont {Lon{\v{c}}ar}},\ }\href@noop {} {\bibfield  {journal}
  {\bibinfo  {journal} {Optica}\ }\textbf {\bibinfo {volume} {7}},\ \bibinfo
  {pages} {1714} (\bibinfo {year} {2020})}\BibitemShut {NoStop}%
\bibitem [{\citenamefont {Bertero}\ \emph {et~al.}(2021)\citenamefont
  {Bertero}, \citenamefont {Boccacci},\ and\ \citenamefont
  {De~Mol}}]{bertero2021introduction}%
  \BibitemOpen
  \bibfield  {author} {\bibinfo {author} {\bibfnamefont {M.}~\bibnamefont
  {Bertero}}, \bibinfo {author} {\bibfnamefont {P.}~\bibnamefont {Boccacci}},\
  and\ \bibinfo {author} {\bibfnamefont {C.}~\bibnamefont {De~Mol}},\
  }\href@noop {} {\emph {\bibinfo {title} {Introduction to inverse problems in
  imaging}}}\ (\bibinfo  {publisher} {CRC press},\ \bibinfo {year}
  {2021})\BibitemShut {NoStop}%
\bibitem [{\citenamefont {Tsang}\ \emph {et~al.}(2016)\citenamefont {Tsang},
  \citenamefont {Nair},\ and\ \citenamefont {Lu}}]{tsang2016quantum}%
  \BibitemOpen
  \bibfield  {author} {\bibinfo {author} {\bibfnamefont {M.}~\bibnamefont
  {Tsang}}, \bibinfo {author} {\bibfnamefont {R.}~\bibnamefont {Nair}},\ and\
  \bibinfo {author} {\bibfnamefont {X.-M.}\ \bibnamefont {Lu}},\ }\href@noop {}
  {\bibfield  {journal} {\bibinfo  {journal} {Phys. Rev. X}\ }\textbf {\bibinfo
  {volume} {6}},\ \bibinfo {pages} {031033} (\bibinfo {year}
  {2016})}\BibitemShut {NoStop}%
\bibitem [{\citenamefont {Zhuang}\ and\ \citenamefont
  {Zhang}(2019)}]{zhuang2019physical}%
  \BibitemOpen
  \bibfield  {author} {\bibinfo {author} {\bibfnamefont {Q.}~\bibnamefont
  {Zhuang}}\ and\ \bibinfo {author} {\bibfnamefont {Z.}~\bibnamefont {Zhang}},\
  }\href {https://doi.org/10.1103/PhysRevX.9.041023} {\bibfield  {journal}
  {\bibinfo  {journal} {Phys. Rev. X}\ }\textbf {\bibinfo {volume} {9}},\
  \bibinfo {pages} {041023} (\bibinfo {year} {2019})}\BibitemShut {NoStop}%
\bibitem [{\citenamefont {Xia}\ \emph {et~al.}(2021)\citenamefont {Xia},
  \citenamefont {Li}, \citenamefont {Zhuang},\ and\ \citenamefont
  {Zhang}}]{xia2021}%
  \BibitemOpen
  \bibfield  {author} {\bibinfo {author} {\bibfnamefont {Y.}~\bibnamefont
  {Xia}}, \bibinfo {author} {\bibfnamefont {W.}~\bibnamefont {Li}}, \bibinfo
  {author} {\bibfnamefont {Q.}~\bibnamefont {Zhuang}},\ and\ \bibinfo {author}
  {\bibfnamefont {Z.}~\bibnamefont {Zhang}},\ }\href
  {https://doi.org/10.1103/PhysRevX.11.021047} {\bibfield  {journal} {\bibinfo
  {journal} {Phys. Rev. X}\ }\textbf {\bibinfo {volume} {11}},\ \bibinfo
  {pages} {021047} (\bibinfo {year} {2021})}\BibitemShut {NoStop}%
\end{thebibliography}
%

\end{document}